\documentclass{elsart}
\usepackage{graphicx,amsmath,amssymb,bm}

\begin{document}

\newcommand{\kf}{k_{\text{F}}}
\newcommand{\be}{\begin{equation}}
\newcommand{\ee}{\end{equation}}
\newcommand{\q}{{\bf q}}
\newcommand{\qp}{{\bf q'}}
\newcommand{\pp}{{\bf p}+{\bf p'}}
\newcommand{\ip}{{\bf p''}}
\newcommand{\la}{\Lambda}
\newcommand{\zf}{$z_{\kf}$}
\newcommand{\spin}{{\bm \sigma} \cdot {\bm \sigma'}}

\begin{frontmatter}
\title{Renormalization group approach to \\neutron matter:
quasiparticle interactions, \\superfluid gaps and the equation of
state}
\author{Achim Schwenk$^{(a)}$\thanksref{AS}},
\author{Bengt Friman$^{(b)}$\thanksref{BF}} and
\author{Gerald E. Brown$^{(a)}$\thanksref{GEB}}
\address{$^{(a)}$Department of Physics and Astronomy,
State University of New York,\\
Stony Brook, N.Y. 11794-3800, U.S.A.\\
$^{(b)}$Gesellschaft f\"ur Schwerionenforschung, Planckstr. 1, 64291
Darmstadt, Germany}

\thanks[AS]{E-mail: aschwenk@nuclear.physics.sunysb.edu}
\thanks[BF]{E-mail: b.friman@gsi.de}
\thanks[GEB]{E-mail: popenoe@nuclear.physics.sunysb.edu}

\begin{abstract}

\noindent
Renormalization group methods can be applied to the nuclear
many-body problem using the approach proposed by Shankar. We start
with the two-body low momentum interaction $V_{\text{low k}}$ and
use the RG flow from the particle-hole channels to calculate the
full scattering amplitude in the vicinity of the Fermi surface.
This is a new straightforward approach to the many-body problem
which is applicable also to condensed matter systems without
long-range interactions, such as liquid $^3$He. We derive the
one-loop renormalization group equations for the quasiparticle
interaction and the scattering amplitude at zero temperature. The RG
presents an elegant method to maintain all momentum scales and
preserve the antisymmetry of the scattering amplitude. As a first
application we solve the RG equations for neutron matter. The
resulting quasiparticle interaction includes effects due to the
polarization of the medium, the so-called induced interaction of
Babu and Brown. We present results for the Fermi liquid parameters, the
equation of state of neutron matter and the $^1$S$_0$ superfluid
pairing gap.

\vspace{0.5cm}

\noindent{\it PACS:}
21.65.+f;          
71.10.Ay;          
11.10.Hi \\        
\noindent{\it Keywords:} Neutron matter; Fermi liquid theory;
Renormalization Group; Polarization Effects; Superfluidity
\end{abstract}
\end{frontmatter}

\section{Introduction}

Fermi liquid theory is a prototype effective theory, invented by
Landau in the late 1950's~\cite{Landau1,Landau2,Landau3}. In this
theory the properties of normal Fermi liquids at zero temperature
are encoded in a few parameters, which are related to the effective
two-fermion interaction. Over the years, Fermi liquid theory has
proven to be an extremely useful tool for studying normal Fermi
liquids, in particular liquid $^3$He~\cite{BaymPethick}. The main
assumption is that the elementary, low-lying excitations of the
interacting system are relatively long-lived quasiparticles, with
properties resembling those of free particles. The remaining
strength in the spectral function is distributed over modes that
add up incoherently. In the language of the renormalization group,
the incoherent background is integrated out into the quasiparticle
interaction, which can be determined by comparison with experiment
or by microscopic calculations.

Babu and Brown later realized~\cite{BB} that, in order to satisfy
the Pauli principle in microscopic calculations, one has to take
into account not only the particle-hole channel considered by
Landau, which gives rise to the propagation of zero sound, but also
the exchange channel thereof. This is taken into account in the
induced interaction, which incorporates the response of the
many-body medium to the presence of the quasiparticles.

Almost a decade ago, Shankar revived the interest in Fermi
liquid theory by developing a renormalization group (RG) approach
to interacting Fermi systems (for an introduction see
Shankar~\cite{FLTandRG1} and the lecture notes of
Polchinski~\cite{FLTandRG2}). Since Fermi liquid theory is
restricted to low-lying excitations in the vicinity of the Fermi
surface, normal Fermi systems are amenable to the renormalization
group, where one has precisely such a separation of modes in mind.
Moreover, by taking the loop contributions from both particle-hole
channels into account, the RG flow remains antisymmetric, i.e., at
any step of the renormalization, the scattering amplitude obeys the
Pauli principle and consequently the solution obeys the Pauli
principle sum rules. Thus, the antisymmetry of the scattering
amplitude, which was the guiding principle of Babu and Brown, is
realized naturally in the RG approach. Recently, we obtained
additional RG invariant constraints~\cite{IIpaper}. These
approximate constraints are analogous to those obtained by Bedell
and Ainsworth for paramagnetic Fermi liquids~\cite{BA}.

In this paper, we solve the one-loop RG equations in the
particle-hole channels at zero temperature and in three dimensions.
We work in the approximation that both particle-hole momentum
transfers are small compared to the Fermi momentum. The
justification for this approximation is two-fold: In Fermi liquid
theory the long-wavelength excitations play a central role. Hence,
our primary aim is to treat these correctly. Furthermore, for the
case of nuclear or neutron matter, the dependence of the induced
interaction on Landau angle is fairly
weak~\cite{BBfornuclmat,bigreport}. Thus, we expect an expansion in
momentum transfers to be accurate at least for the Fermi liquid
parameters.\footnote{This may change when effects of the tensor
interaction are included in the induced interaction.} In this
approximation the treatment of the particle-hole phase space with
cutoffs is enormously simplified. An RG analysis of a schematic
model, including the complete particle-hole phase space in two
dimensions, is discussed in~\cite{BF}. Improved RG equations, beyond one-loop
order, are presented in~\cite{IIpaper}.

Here we apply the RG techniques to neutron matter, where the tensor force
does not contribute in the S-wave. In this system, a further
physical motivation for expanding in small momentum transfers over
the Fermi momentum is provided by the unique low momentum
nucleon-nucleon interaction $V_{\text{low
k}}$~\cite{Vlowk,Vlowkflow}, which we employ as the starting point
for the RG flow. As shown in Figs.~2 and~4 of Bogner {\em et
al.}~\cite{Vlowkflow}, $V_{\text{low k}}$ is significant only for
relative momenta $k < 1.3\,\text{fm}^{-1}$. Moreover, from the
evolution of $V_{\text{low k}}$ shown in Fig.~5
of~\cite{Vlowkflow} we deduce that the important momentum modes
are around the pion mass $m_\pi \approx 0.7\,\text{fm}^{-1}$, since
higher momenta do not renormalize $V_{\text{low k}}$ considerably.
Therefore, the typical momentum transfer is small compared to the
Fermi momentum of neutron matter at nuclear matter density $\kf =
1.7\,\text{fm}^{-1}$.

The main idea of the RG approach to the many-body problem is to
``adiabatically'' include the in medium corrections to the
effective interaction by solving the RG flow equations in the
relevant channels. In this exploratory calculation we include the
particle-hole channels, which play a special role in Fermi liquid
theory. The main effect of scattering in the particle-particle
channel, the removal of the short-range repulsion, is taken care
of by using $V_{\text{low k}}$ as the starting point for the RG
flow. The low-lying excitations in this channel, which are
responsible e.g., for superfluidity, are not included. These are
then treated explicitly in Section~\ref{gapsec}, where we compute the
superfluid gap by employing BCS theory for the particle-hole reducible
scattering amplitude.

In the RG approach, the width of the momentum shell that is
integrated out in one iteration is a ``small parameter''. The full
scattering amplitude is an RG invariant quantity. In a conventional
approach, this is calculated from a ``set of diagrams'' while in
the RG approach it is continuously evolved from the starting
interaction by gradually including the many-body corrections from
the narrow momentum shells one at a time. In this sense the RG
provides a method for dealing with strongly interacting systems,
where perturbation theory fails. We stress that this application of
the renormalization group to many-body systems is only indirectly
related to the well known RG treatment of critical phenomena.

We derive the one-loop renormalization group equations for the
quasiparticle interaction and the scattering amplitude at zero
temperature. The evolution of the effective mass is included in the
RG flow, as well as a simplified treatment of the renormalization
of the quasiparticle strength. In the forward scattering limit, the
role of the energy transfer is in the RG taken over by the cutoff
in momentum space. We find that, in the long-wavelength limit, the
dependence of the effective four-point vertex on $\Lambda/q$
corresponds to the behavior with $\omega/q$ in the microscopic
derivation of Fermi liquid theory by Landau.

Finally, we present the solution of the RG equations for neutron
matter. These are obtained by employing the unique low momentum
nucleon-nucleon interaction as the initial condition of the RG. Our
results include the Fermi liquid parameters in the density range of
interest for neutron stars as well as the full scattering amplitude
for general (non-forward) scattering on the Fermi surface. Using
the Fermi liquid parameters, we compute the equation of state,
including polarization effects, by integrating the
incompressibility~\cite{Kaellman}. Finally, we compute the
$^1$S$_0$ superfluid pairing gap using weak coupling BCS theory.
This is an application of our approach, which probes the
angular dependence of the scattering amplitude. We generally find
very good agreement with the results obtained in the polarization
potential model by Ainsworth, Wambach and Pines~\cite{WAP1,WAP2}.
However, it is worth noting that much of the model dependence,
inherent in the work of Ainsworth {\em et al.}, can be avoided in
the RG approach.

\section{Renormalization group at one-loop}

Shankar suggested that the relevant modes of a normal Fermi system
may be isolated in a similar manner as for critical phenomena. In
analogy with a Wilson-Kadanoff treatment of the
latter~\cite{Wilson}, he proposed to separate the ``slow'' modes
from the ``fast''\footnote{With ``slow'' we refer to the modes
within a shell of width $2 \Lambda$ centered around the Fermi
momentum $\kf$, whereas the ``fast'' modes lie outside this shell.}
ones by imposing a cutoff around the Fermi surface~\cite{Shankar}.
Using a loop expansion, one can systematically generate the
effective quasiparticle scattering amplitude and the effective
quasiparticle interaction among slow quasiparticles and slow
quasiholes at a given scale $\Lambda$.\footnote{The effective
scattering amplitude at the scale $\Lambda$ interpolates between
the two-body irreducible interaction and the full scattering amplitude
as described in the introduction.} It has been shown that, in this
approach, Fermi liquid theory emerges in the infrared limit when
the cutoff is taken to
zero~\cite{FLTandRG1,FLTandRG2,Shankar,Dupuis,Benfatto1,Benfatto2,Feldman,DuChi,Hewson}.

Recently we derived the flow equations for the effective scattering
amplitude and the quasiparticle interaction from the induced
interaction of Babu and Brown~\cite{IIpaper}. In this paper, we
employ the one-loop RG equations to compute the full scattering
amplitude and the quasiparticle interaction on the Fermi surface
starting from the vacuum two-body interaction. As we decimate down
to the Fermi surface by letting $\la \to 0$, we obtain not only the
forward scattering amplitude for low-lying quasiparticle-quasihole
excitations, but also the amplitude for general (non-forward)
scattering processes of quasiparticles on the Fermi surface in the
small momentum approximation.

In a loop expansion, the induced interaction corresponds to a
summation of the direct and exchange particle-hole channels in a
particular kinematic window, while loop contributions from the
particle-particle (BCS) channel, are not summed by the RG equations
derived in~\cite{IIpaper}. Thus, the contributions from the BCS
channel should be included in the particle-hole irreducible driving
term. In practice, one usually restricts oneself to the two-particle
ladders with the bare nucleon-nucleon interaction (the Brueckner $G$
matrix) or a simple generalization thereof, where the lowest order
renormalization of the quasiparticle strength is
included~\cite{BBfornuclmat}.

In the present work, we explore the one-loop truncation of the RG
equations. When only one channel is considered, the one-loop RG
equation is exact in the sense that it is equivalent to the
corresponding scattering equation. When both the direct and
exchange channels are included, the scattering amplitude remains
antisymmetric under the RG flow. Calculations with the induced
interaction for liquid $^3$He and nuclear matter show that the
systematic inclusion of exchange channel contributions is crucial
for obtaining a realistic description of the effective
interaction~\cite{BA,BBfornuclmat,bigreport,tensor,FLnm,He3.1,He3.2}.
Therefore, we expect the one-loop RG to capture the essential
physics of the soft modes near the Fermi surface. In this
approximation, our RG equations are equivalent to those obtained
previously by Dupuis~\cite{Dupuis}.

\begin{figure}[t]
\begin{center}
\includegraphics[scale=0.9,clip=]{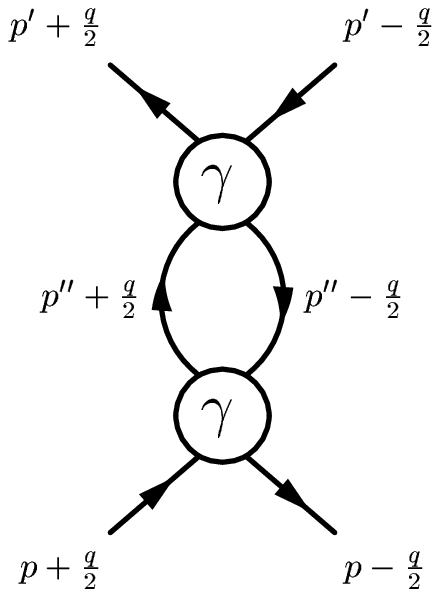}
\hspace{1cm}
\includegraphics[scale=0.9,clip=]{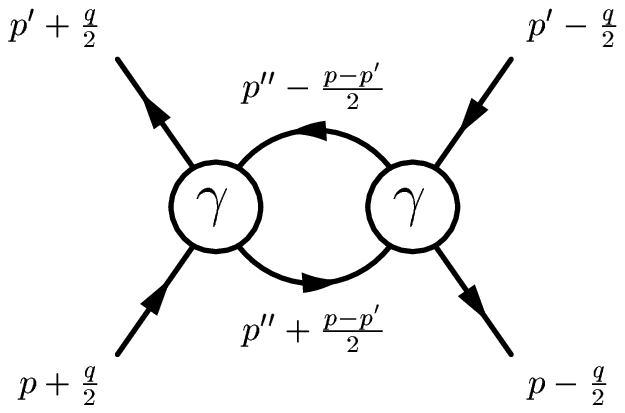}
\end{center}
\hspace{3.1cm} (A) \hspace{5.15cm} (B)
\caption{The one-loop contributions to the RG equation.
(A): The zero sound (ZS) channel with momentum transfer ${\bf q}$
and (B): the exchange channel (ZS') with momentum transfer ${\bf
q'}$. (Our convention for the momentum flow is in the direction of
the particle and hole arrows.) As defined in the text, $\gamma$
denotes the effective four-point vertex.}
\label{zsandzsp}
\end{figure}

The two one-loop particle-hole contributions to the effective
four-point function are shown in Fig.~\ref{zsandzsp}, where the
integration is restricted to a fast particle and a fast hole. The
effective four-point vertex $\gamma$ depends on three momentum
variables, the cutoff and the energy transfer $\omega$ in the zero
sound channel,\footnote{For energy-independent initial conditions,
one can set the energy transfers $\omega$ and $\omega^\prime$ to
zero in the RG equations. Here we retain only the $\omega$ for the
purpose of discussing the limits $\omega/q\to 0$ and $q/\omega\to
0$. The role of $\omega$ for the purpose of obtaining the forward
scattering amplitude and the quasiparticle interaction is then
taken over by the cutoff $\la$, as we will show in the next
section.}
\begin{equation}
\gamma = \gamma((\omega,\q),\qp,\pp;\la) ,
\end{equation}
where ${\bf q'} = {\bf p} - {\bf p'}$. Under exchange, the momentum
transfer $\q$ and $\qp$ have to be interchanged, whereas the
particle-pair momentum $\pp$ remains invariant. We note that in the
limit $|\q|\to 0$, the scattering amplitude describes
long-wavelength excitations in the zero sound channel, while
$|\pp|\to 0$ corresponds to the BCS pairing singularity.

We define the effective quasiparticle interaction $f(\qp;\la)$ at
the scale $\la$ by the limit
\begin{equation}
f(\qp;\la) = \lim_{\omega \to 0}
\gamma((\omega,\q=0),\qp,\pp;\la) ,
\label{efff}
\end{equation}
and the effective quasiparticle scattering amplitude,
extended to finite momentum transfer $\q$, by
\begin{equation}
a(\q,\qp;\la) = \gamma((\omega=0,\q),\qp,\pp;\la) .
\label{effa}
\end{equation}
In order to allow a proper treatment of the momentum dependence in
the exchange channel, we generalize the effective interaction
$f(\qp;\la)$ to finite $\q$, $f(\q,\qp;\la)$, keeping in mind that
the Landau quasiparticle interaction is obtained for $|\q| = 0$.

The RG equations at one-loop and zero temperature are given
by~\cite{IIpaper}
\begin{align}
\frac{d}{d \Lambda} a(\q,\qp;\la) &= z_{\kf}^2 \:
\frac{d}{d \Lambda} \biggl\{ \;
g \int\limits_{\text{fast}, \Lambda} \frac{d^3 \ip}{(2 \pi)^3} \:
\frac{n_{\ip+\q/2}-n_{\ip-\q/2}}{\varepsilon_{\ip+\q/2}-\varepsilon_{\ip-\q/2}}
\biggr\} \nonumber \\[1mm]
& \times \; a\bigl(\q,\frac{\pp}{2}+\frac{\qp}{2}-\ip;\la\bigr) \;
a\bigl(\q,\ip-\frac{\pp}{2}+\frac{\qp}{2};\la\bigr) \nonumber
\\[1mm]
& + \frac{d}{d \Lambda} f(\q,\qp;\la) \label{arg} \\[2mm]
\frac{d}{d \Lambda} f(\q,\qp;\la) &= - \, z_{\kf}^2 \:
\frac{d}{d \Lambda} \biggl\{ \;
g \int\limits_{\text{fast}, \Lambda} \frac{d^3 \ip}{(2 \pi)^3} \:
\frac{n_{\ip+\qp/2}-n_{\ip-\qp/2}}{\varepsilon_{\ip+\qp/2}
-\varepsilon_{\ip-\qp/2}}
\biggr\} \nonumber \\[1mm]
& \times \; a\bigl(\qp,\frac{\pp}{2}+\frac{\q}{2}-\ip;\la\bigr) \;
a\bigl(\qp,\ip-\frac{\pp}{2}+\frac{\q}{2};\la\bigr) ,
\label{frg}
\end{align}
where $g$ denotes the spin-isospin degeneracy, for neutron matter
$g=2$, and \zf~is the quasiparticle strength at the Fermi surface.
The effective scattering amplitude includes contributions
from both the ZS and ZS' channel, whereas for the effective
interaction only the exchange contribution, which corresponds to
the induced interaction~\cite{BB}, remains. The vertices in the
one-loop diagrams are given by the scattering amplitude at the
current scale. Consequently, the renormalization due to the fast modes in
both particle-hole channels is included. Furthermore, since we will
employ the unique, energy-independent low momentum nucleon-nucleon
interaction of Bogner {\em et al.}~\cite{Vlowk,Vlowkflow} as initial
condition of the RG flow, we have set the energy transfer to zero,
in particular $\omega=0$, in the above RG equations. Note that the
derivatives with respect to $\Lambda$ on the right hand side of
Eqs.~(\ref{arg},\ref{frg}) act only on the phase space factors.
Furthermore, in order not to overload the equations, we have
suppressed the spin variables. They will be discussed in detail
below. It is apparent that the RG equation, Eq.~(\ref{arg}),
preserves the antisymmetry of the scattering amplitude under the RG
flow. Hence, the Pauli principle sum rules~\cite{Landau3,BD} are
automatically fulfilled for the resulting Fermi liquid parameters.

\begin{figure}[t]
\begin{center}
\includegraphics[scale=0.35,clip=]{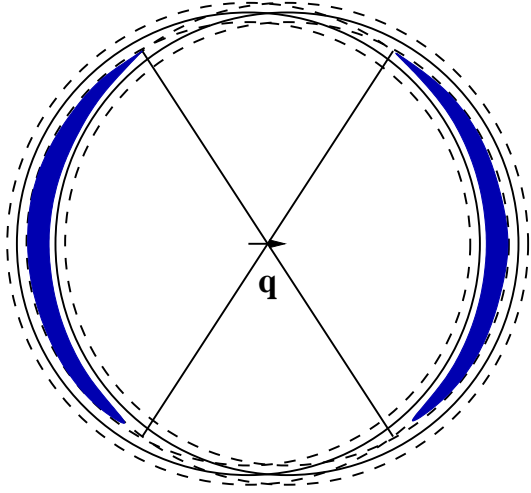}
\end{center}
\caption{The phase-space factors in the RG equation,
Eq.~(\ref{arg}), are non-zero in the filled areas. The dashed
circles denote the cutoff around the Fermi surfaces.}
\label{conf}
\end{figure}
As motivated in the introduction, we evaluate the one-loop beta
functions in the long-wavelength approximation. Obviously, the phase
space for the intermediate particle-hole pair in the ZS channel is
open for $\la < q/2$, where $q$ denotes $|\q|$, with an analogous
expression for the ZS' channel, see Fig.~\ref{conf}. For small $q, q'$, the
integration over $\ip$ in the beta functions reduces to an
integration over a solid angle on the Fermi surface. We find for
the RG equations at zero temperature\footnote{Contrary to
statements in the literature, the RG equations at zero temperature
are very transparent. The $\delta$ functions appearing in the beta
function can be handled in a straightforward manner and the phase
space integrals can in fact be evaluated analytically. A non-zero
temperature introduces a further scale, which competes with the
cutoff and consequently must be treated with special care.}
\begin{align}
\Lambda\,\frac{d}{d \Lambda} a(\q,\qp;\la) &= \Theta(q-2\la) \:
\beta_{\text{ZS}}[a,\q,\la] - \Theta(q'-2\la) \:
\beta_{\text{ZS'}}[a,\qp,\la] \label{thetabeta} \\[1mm]
& = \Theta(q-2\la) \: z_{\kf}^2 \: \frac{2\Lambda}{q} \: N(\la)
\nonumber \\[1mm] 
& \times \int_{2\la/q<|\cos(\theta_{\ip\q})|<1} \hspace{-2mm} \frac{d
\Omega_{\ip}}{4 \pi} \,
\frac{\text{sign}\bigl(\cos(\theta_{\ip\q})\bigr)}{\cos(\theta_{\ip\q})}
\nonumber \\[1mm]
& \times a\bigl(\q,\frac{\pp}{2}+\frac{\qp}{2}-\ip;\la\bigr) \;
a\bigl(\q,\ip-\frac{\pp}{2}+\frac{\qp}{2};\la\bigr) \nonumber
\\[1mm]
& - \Theta(q'-2\la) \: z_{\kf}^2 \: \frac{2\Lambda}{q'} \:
\bigl\{ \q \leftrightarrow \qp \bigr\} \label{rgeq} \\[2mm]
\Lambda\,\frac{d}{d \Lambda} f(\q,\qp;\la) &= - \Theta(q'-2\la) \: z_{\kf}^2 \:
\frac{2\Lambda}{q'} \: N(\la) \nonumber \\[1mm]
& \times \int_{2\la/q'<|\cos(\theta_{\ip\qp})|<1} \hspace{-2mm} \frac{d
\Omega_{\ip}}{4 \pi} \,
\frac{\text{sign}\bigl(\cos(\theta_{\ip\qp})\bigr)}{\cos(\theta_{\ip\qp})}
\nonumber \\[1mm]
& \times a\bigl(\qp,\frac{\pp}{2}+\frac{\q}{2}-\ip;\la\bigr) \;
a\bigl(\qp,\ip-\frac{\pp}{2}+\frac{\q}{2};\la\bigr) .
\label{fofrgeq}
\end{align}
The factor $N(\la)=g \, m^\star(\la) \, \kf / \, 2 \, \pi^2$
denotes the density of states at the Fermi surface. The effective
mass is determined self-consistently from the effective interaction
\begin{equation}
\frac{m^\star(\la)}{m} = \cfrac{1}{1 - f_1(\la) \: \cfrac{g \,
m \, \kf}{6 \, \pi^2}}\,.
\label{mstar}
\end{equation}
This enforces Galilean invariance on the system.
The effective mass is continuously adapted as the flow equations
are integrated towards the Fermi surface. In Eq.~(\ref{mstar}),
$f_1(\la)$ is given by the projection of the quasiparticle
interaction on the $l=1$ Legendre polynomial. If one restricts the
analysis to the ZS channel and assumes that the Fermi liquid
parameters are given, the solution of the flow equation reproduces
the relations between the quasiparticle interaction and the forward
scattering amplitude. In the appendix we give a detailed derivation
of the beta function in Eqs.~(\ref{thetabeta}-\ref{fofrgeq}) and the
solution of the RG equation in the ZS channel.

For $\la \geqslant \kf$, the particle-hole contributions vanish in
the momentum range of interest $q,q' \leqslant 2 \, \kf$ (see the
explicit expressions given in the appendix). Hence, we can start
the decimation at $\la_0 = \kf$ with a particle-hole irreducible
boundary condition on the RG equations,
Eqs.~(\ref{rgeq},\ref{fofrgeq}). This boundary condition may be
identified with the direct interaction of~\cite{BB}. We approximate
the direct interaction with the vacuum two-body low momentum
interaction $V_{\text{low k}}$ at a cutoff scale $\la_{V_{\text{low
k}}}=\sqrt{2} \: \kf$, which will be motivated in Section~\ref{results}.
This choice of the cutoff approximately accounts for the Pauli
blocking of intermediate states in the BCS channel, which
suppresses large structures, like the quasi bound state in the
$^1$S$_0$ channel and tames the short-range repulsion of the
nucleon-nucleon interaction. Since, in isotriplet channels,
$V_{\text{low k}}$ is almost independent of the cutoff over a large
range, the precise value of $\la_{V_{\text{low k}}}$ is not
crucial. Furthermore, we allow for a minimal in medium correction
of the starting values, in terms of the renormalization of the
quasiparticle strength $z_{\kf}$. To a certain extent this accounts
for the contribution of higher order particle-hole irreducible
diagrams to the direct interaction~\cite{BBfornuclmat}. At the
initial cutoff $\la_0$, we thus have
\begin{equation}
a(\q,\qp;\la_0) = f(\q,\qp;\la_0) = \gamma_{\text{vacuum}}(\q,\qp) ,
\end{equation}
where $\gamma_{\text{vacuum}}(\q,\qp)$ will be specified below. We
solve the RG flow with two different assumptions for the
\zf~factor. In one case, we will use a static,
density-independent mean value of $z_{\kf}^2 = 0.9$, which remains
unchanged under the RG. In the other case, we compute the
\zf~factor dynamically, by assuming that the change of the
effective mass from the initial one, owing to the direct
interaction, is due to the \zf~factor alone. We then adjust 
the factor $z_{\kf}^2$ in the RG equations,
Eqs.~(\ref{rgeq},\ref{fofrgeq}), self-consistently as we decimate
to the Fermi surface, i.e., we start with $z_{\kf}(\la_0) = 1$ and
change the \zf~factor according to
\begin{equation}
\frac{1}{z_{\kf}(\la)} = \frac{m^\star(\la)}{m^\star(\la_0)} .
\label{zfactrel}
\end{equation}
In this approximation we account for the momentum dependence of the
self-energy corresponding to the direct interaction in
$m^\star(\la_0)$ and make the reasonable assumption that the
modification of the effective mass due to the induced interaction
is dominated by the energy dependence of the self energy, the
so called $E$-mass~\cite{Mahaux}. We shall see later that this
approximation leads to a
\zf~factor in rather good agreement with Brueckner-Hartree-Fock
calculations~\cite{Zuo,BG} at intermediate Fermi momenta $0.5 \,
\text{fm}^{-1} < \kf < 1.5 \, \text{fm}^{-1}$.

\begin{figure}[t]
\begin{center}
\includegraphics[scale=0.4,clip=]{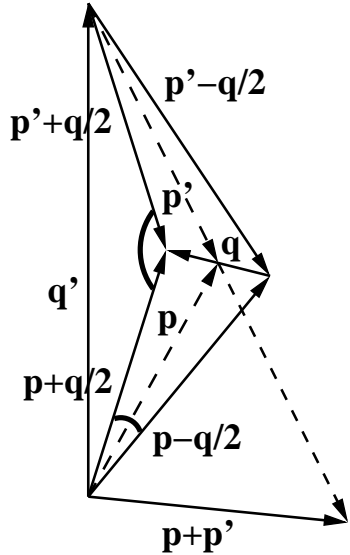}
\end{center}
\caption{The configuration of quasiparticles and quasiholes on the
Fermi surface. The arcs denote the angles $\theta_\q$ and
$\theta_\qp$ opposite of $\q$ and $\qp$, respectively.}
\label{cosys}
\end{figure}
Finally, we assume that the interacting particles are on the Fermi
surface. The momentum variables $\q,\qp$ and $\pp$ then form an
orthogonal set of basis vectors, which follows straightforwardly
from the conditions $|{\bf p} \pm \q/2| = \kf$ and $|{\bf p'} \pm
\q/2| = \kf$. Our geometry is shown in Fig.~\ref{cosys}, where
$\theta_\q$ and $\theta_\qp$ denote the angles between ${\bf p} +
\q/2$, ${\bf p} - \q/2$ and ${\bf p} + \q/2$, ${\bf p'} +
\q/2$ respectively and are related to the momentum transfer by means
of\footnote{Our definition of $\theta_\qp$ differs from the angle
$\theta_{\text{L}}$ between the momenta ${\bf p}$ and ${\bf p'}$
employed by Ainsworth and Bedell~\cite{He3.2}. The relation between
the two angles is given by $2 \cos\theta_\qp = 1 - \cos\theta_\q +
(1 + \cos\theta_\q) \cos\theta_{\text{L}}$.}
\begin{align}
q &= 2 \, \kf \sin \frac{\theta_\q}{2} \\[1mm]
q' &= 2 \, \kf \sin \frac{\theta_\qp}{2} \\[1mm]
|\pp| &= 2 \, \kf \: \sqrt{\frac{\cos\theta_\q + \cos\theta_\qp}{2}} .
\label{ppexact}
\end{align}
The orthogonality of the coordinate system is very convenient
for the numerical implementation, in particular for the evaluating
the arguments of the scattering amplitude in the beta functions.
Given this geometry, the scattering amplitude depends only on the
magnitude of the two momentum transfers $q$ and $q^\prime$. Under
the restriction that the momenta of the quasiparticles are
constrained to the Fermi surface, we have $q^2 + q^{\prime
2}\leqslant 4\,\kf^2$ or equivalently 
$\cos\theta_\q + \cos\theta_\qp \geqslant 0$.

The beta functions involve scattering to fast particle-hole
intermediate states away from the Fermi surface. In the coordinate
system shown in Fig.~\ref{cosys} one finds
\begin{align}
{\bf p} &= \kf \cos \frac{\theta_\q}{2} \: \widehat{\bf p} \\[1mm]
{\bf p'} &= \kf \cos \frac{\theta_\q}{2} \: \widehat{\bf p}^\prime .
\end{align}
Using these relations we obtain, for small $q$, where $\widehat{\bf p}
\cdot \widehat{\bf p}^\prime \approx \cos\theta_\qp$, the
approximate expression
\begin{equation}
|\pp| = 2 \, \kf \, \cos\frac{\theta_\q}{2} \,
\cos\frac{\theta_\qp}{2} ,
\label{ppapprox}
\end{equation}
which allows us to smoothly extend the RG flow to $q^2 + q^{\prime
2} > 4 \, \kf^2$, where Eq.~(\ref{ppexact}) is not
useful.

\section{The forward scattering limit in the RG and the inclusion of spin}

At the one-loop level, the solution of the RG flow in the
particle-hole channels is particularly transparent and involves
only the effective scattering amplitude. The quasiparticle
interaction at arbitrary cutoff is simply given by the $q=0$ values
of the scattering amplitude,
\begin{equation}
\frac{d}{d \la} f(q=0,\qp;\la) = \frac{d}{d \la} a(q=0,\qp;\la) .
\end{equation}

In Fermi liquid theory, the singularity in the ZS channel can be
used to eliminate the quasiparticle-quasihole contributions in this
channel by taking the limit $q/\omega \to 0$. For an
energy-independent RG flow, the role of $\omega$ is taken over by the
cutoff. In the RG equation, Eq.~(\ref{rgeq}), the ZS contribution
can be explicitly eliminated by setting $q$ to zero (see also
Fig.~\ref{conf}, where the contributions from a fast particle and a
fast hole are absent for $q=0$). Therefore, the forward scattering
limits in Fermi Liquid Theory (FLT) are in the RG approach at zero
temperature obtained by
\begin{align}
f_{\text{FLT}}(\qp) & = \lim_{\la \to 0} a(q=0,\qp;\la)
\label{fLandau} \\[1mm]
a_{\text{FLT}}(\qp) & = \lim_{q \to 0} a(\q,\qp;\la=0) .
\label{aLandau}
\end{align}
This is analogous to the $q/\omega\to 0$ and $\omega/q\to 0$ limits
in Fermi liquid theory. The forward scattering amplitude and the
Fermi liquid quasiparticle interaction are obtained as stable
solutions of the RG equation, Eq.~(\ref{rgeq}), as the cutoff is
taken to zero. We note that the beta functions explicitly depend on
the cutoff and the momentum transfer in the particular channel. In
fact, $\displaystyle \la \, \frac{d}{d \la} a(\q,\qp;\la)$ is
proportional to the
cutoff itself, and thus the beta functions vanish by construction as
the cutoff is taken to zero. From the definition of the
quasiparticle interaction, Eq.~(\ref{fLandau}), we observe that the
Fermi liquid parameters depend in a nontrivial fashion on the
effective scattering amplitude.

The inclusion of spin is straightforward. The RG equation for the
scattering amplitude, $a(\q,\qp;\la) + b(\q,\qp;\la) \: \spin$, 
is given by
\begin{align}
\Lambda\,\frac{d}{d \Lambda} \bigl( a + b \: \spin \bigr) &= \Theta(q-2\la) \:
\bigl\{ \beta_{\text{ZS}}[a,\q,\la] + \beta_{\text{ZS}}[b,\q,\la] \: \spin
\bigr\} \nonumber \\[1mm]
& - \Theta(q'-2\la) \: \bigl\{ \frac{1}{2} \,
\beta_{\text{ZS'}}[a,\qp,\la] + \frac{3}{2} \, \beta_{\text{ZS'}}[b,\qp,\la]
\bigr\} \nonumber \\[1mm]
& - \Theta(q'-2\la) \: \bigl\{ \frac{1}{2} \,
\beta_{\text{ZS'}}[a,\qp,\la] - \frac{1}{2} \, \beta_{\text{ZS'}}[b,\qp,\la]
\bigr\} \: \spin ,
\label{rgeqspin}
\end{align}
where the coefficients in Eq.~(\ref{rgeqspin}) are the recoupling
coefficients generated by the spin exchange operator
(see~\cite{IIpaper}). In~\cite{RG2002}, the RG flow is
illustrated for a schematic model of spin polarized fermions.

\section{Results}
\label{results}

We initialize the RG flow with the unique low momentum
nucleon-nucleon interaction $V_{\text{low k}}$ of Bogner {\em et
al.}~\cite{Vlowk,Vlowkflow}. This is derived from realistic
nucleon-nucleon potentials, such as the Bonn, Paris, Argonne and
Nijmegen interactions, by integrating out the high momentum modes
while preserving the half-on-shell $T$ matrix and consequently the
experimental phase shifts. In $V_{\text{low k}}$ the short-range
repulsion has been tamed. Therefore, it is possible to perform
many-body calculations directly with $V_{\text{low k}}$ without
first computing the Brueckner $G$ matrix. In fact, for
$\la_{V_{\text{low k}}} \sim \kf$, $V_{\text{low k}}$ exhibits many
of the salient features of the $G$ matrix~\cite{IIpaper}.

In the case of a static, cutoff independent \zf~factor, a minimal
in medium correction to the direct interaction is, as discussed
above, included by multiplying $V_{\text{low k}}$ by $z_{\kf}^2 $
from the start of the RG. After projecting out the non-central
components, the direct and exchange contribution are in a partial
wave basis given by
\begin{align}
\bigl\langle \, S \, \bigl| \, \mathcal{A}(\q,\qp;\la_0) \, \bigr| \,
S \, \bigr\rangle
&= z_{\kf}^2(\la_0) \: N(\la_0) \: \frac{4 \pi}{2 S + 1} \: \sum_{J \: ,
\: l} \: ( 2 J + 1 ) \: \bigl( 1 - (-1)^{l+S+1} \bigr) \nonumber \\[1mm]
&\times \biggl\langle \, \frac{|\qp-\q|}{2} \: l \: S \: J \: T=1 \, \biggl|
\, V_{\text{low k}} \, \biggr| \, \frac{|\qp+\q|}{2} \: l \: S \: J \: T=1
\, \biggr\rangle \nonumber \\[1mm]
&\times P_l\biggl( \frac{(\qp+\q) \cdot (\qp-\q)}{|\qp+\q| \,
|\qp-\q|} \biggr) ,
\label{direct}
\end{align}
where we set $T=1$ for neutron matter and $P_l$ denotes the
Legendre polynomial of degree $l$. We include all partial waves up
to $J \leqslant 6$ in the driving term. Furthermore, we have introduced
the dimensionless scattering amplitude $\mathcal{A}$ by means of
\begin{align}
\mathcal{A}(\q,\qp;\la) &= A(\q,\qp;\la) + B(\q,\qp;\la) \: \spin
\\[1mm]
&= z_{\kf}^2(\la) \: N(\la) \: \bigl( \, a(\q,\qp;\la) + b(\q,\qp;\la)
\: \spin \, \bigr) . \nonumber
\end{align}

\begin{figure}[t]
\begin{center}
\includegraphics[scale=0.34,clip=]{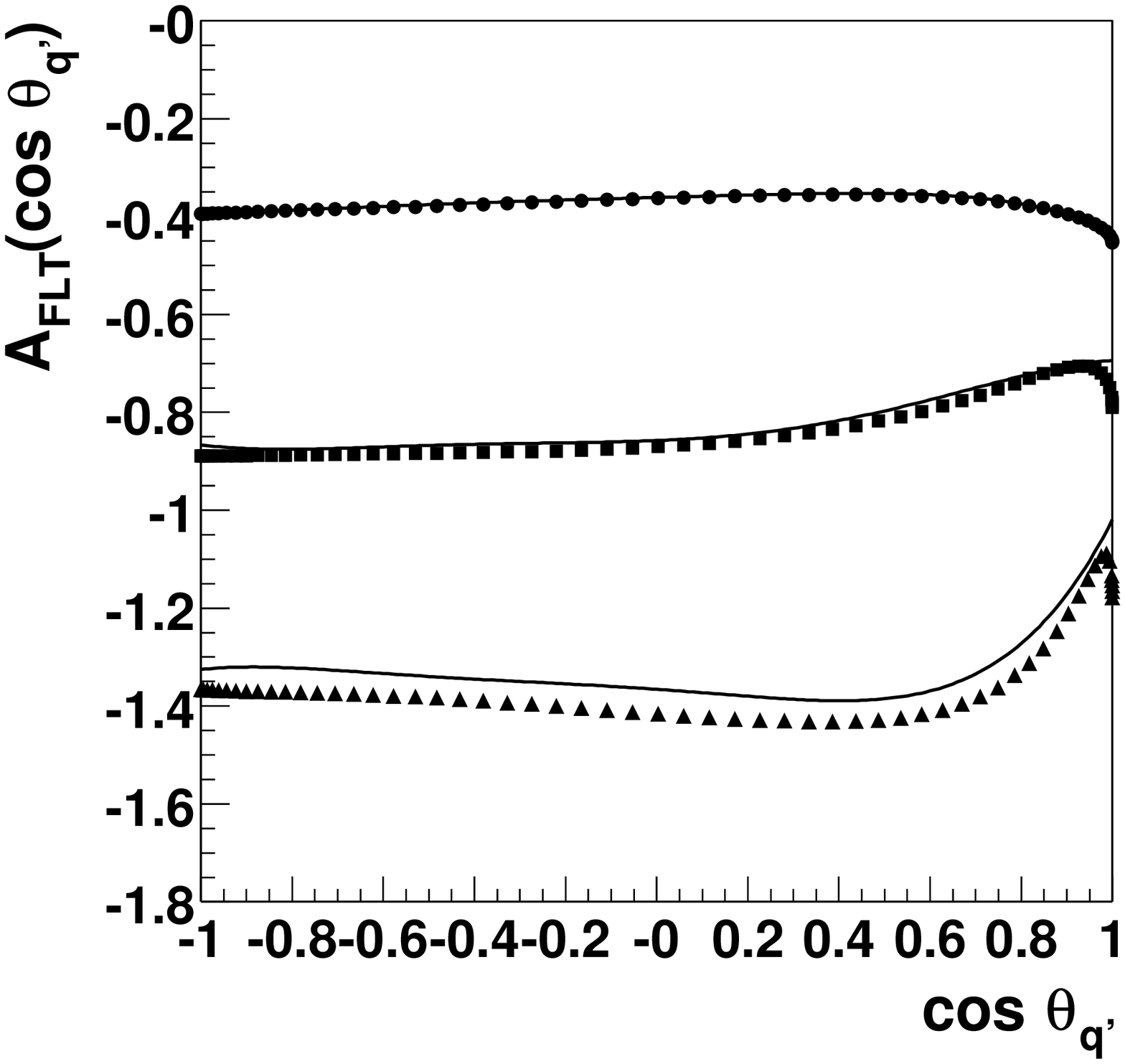}
\includegraphics[scale=0.34,clip=]{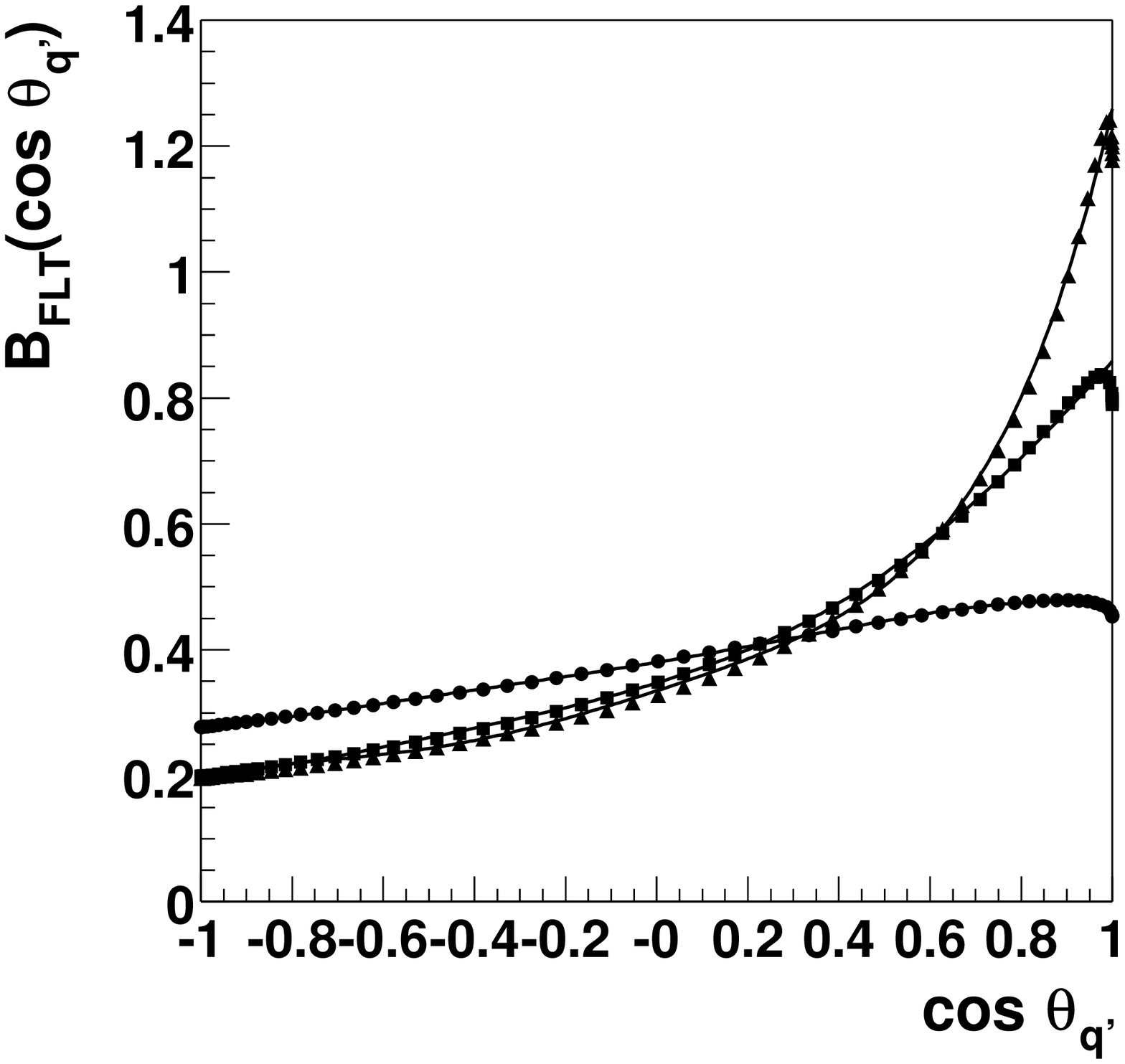}
\end{center}
\caption{The comparison of the RG solution for the forward scattering
amplitude $A_{\text{FLT}}(\cos\theta_\qp)$ and
$B_{\text{FLT}}(\cos\theta_\qp)$ (points) to that obtained from the
corresponding quasiparticle interaction (The solid lines include
the Fermi liquid parameters up to $l \leqslant 5$.). The solutions given
are for the adaptive \zf~factor and a density corresponding to
$\kf= 0.6\, \text{fm}^{-1}$ (dots), $\kf = 1.0 \,
\text{fm}^{-1}$ (squares) and $\kf = 1.35 \,
\text{fm}^{-1}$ (triangles). Similar agreement is found at
other densities. The average momentum grid spacing is $\Delta q/\kf =
\Delta q'/\kf = 0.033$.}
\label{farel}
\end{figure}
In the scattering geometry discussed in the previous section, $\q$
and $\qp$ are orthogonal, and thus the magnitude of the relative
momentum in the initial and final states are both given by
\begin{equation}
\frac{|\qp \pm \q|}{2} = \frac{\sqrt{q^{\prime 2}+q^2}}{2} .
\end{equation}
Let us now discuss the choice of the cutoff in the direct
interaction, $\la_{V_{\text{low k}}}$. In~\cite{IIpaper} we argued
that for the scattering of two particles with total momentum $\pp =
0$ relative to the Fermi sea, the
choice $\la_{V_{\text{low k}}}=\kf$ accounts for the Pauli blocking
of the intermediate states. For non-zero total momenta, the effect
of Pauli blocking is on the average more severe. The Pauli blocking
factor, averaged over the total momentum of two particles on the
Fermi surface, corresponds to an angular averaged cutoff of $\simeq
1.5 \: \kf$. We choose the cutoff for the low momentum
nucleon-nucleon interaction at $\la_{V_{\text{low k}}}
=\sqrt{2} \: \kf$.\footnote{Since the cutoff dependence
of $V_{\text{low k}}$ is very weak for $\la >
0.7$ fm$^{-1}$~\cite{Vlowkflow}, the exact value
of $\la_{V_{\text{low k}}}$ is not decisive.}
This choice has two additional somewhat technical advantages. First,
inspection of the flow equations, Eqs.~(\ref{arg},\ref{frg}), reveals 
that momentum transfers up to $k_{\text{max}}=\sqrt{5}\,\kf / 2$ are needed to
scatter to intermediate states away from the Fermi surface.
Since $V_{\text{low k}}$ is defined only for momenta less than the
cutoff, one would like to have $k_{\text{max}}<\la_{V_{\text{low k}}}$ in
a consistent description. Second, a cutoff $\la_{V_{\text{low k}}}
=\sqrt{2} \: \kf$ allows one to define the scattering amplitude
continuously on the momentum grid $q,q' \leqslant 2 \, \kf$.

\begin{figure}[t]
\begin{center}
\includegraphics[scale=0.35,clip=]{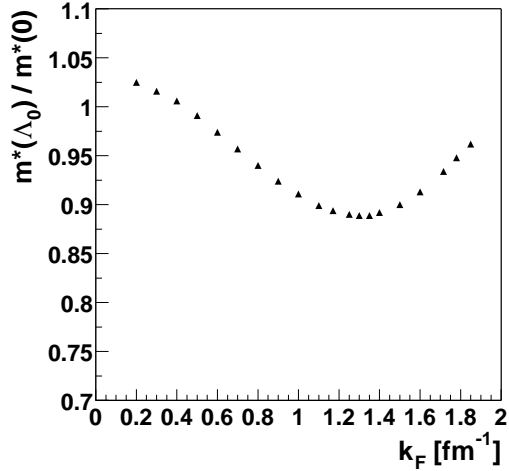}
\end{center}
\caption{The ratio of the direct to the full effective mass
$m^\star(\la_0)/m^\star(\la=0)$, which we use as a running \zf~factor in
the RG equation, versus Fermi momentum $\kf$.}
\label{zfact}
\end{figure}
The numerical procedure for solving the RG equations,
Eq.~(\ref{rgeq}) or Eq.~(\ref{rgeqspin}), may be checked in the
following way. First, the antisymmetry of the effective scattering
amplitude is by construction preserved under the RG
\begin{equation}
\frac{d}{d \Lambda} \: \bigl\langle \, S=1 \, \bigl| \, 
\mathcal{A}(\q,\qp;\la) + \mathcal{A}(\qp,\q;\la) \, \bigr| \, S=1 \,
\bigr\rangle = 0 .
\end{equation}
Furthermore, the Fermi liquid quasiparticle interaction and the
forward scattering amplitude, Eqs.~(\ref{fLandau},\ref{aLandau}),
must satisfy general relations of Fermi liquid theory. The Fermi
liquid parameters are given by the projection of the quasiparticle
interaction on Legendre polynomials
\begin{equation}
\label{legendre}
\mathcal{F}_{\text{FLT}}(\cos\theta_\qp) = z_{\kf}^2(0) \: N(0)
f_{\text{FLT}}(\qp) = \sum_l \bigl( F_l + G_l \:
\spin \bigr) \: P_l(\cos\theta_\qp) ,
\end{equation}
where $\theta_\qp$ is the angle between the quasiparticle momenta
${\bf p}$ and ${\bf p'}$ in the limit $q = 0$. Similarly, the
forward scattering amplitude is expanded in Legendre polynomials
\begin{equation}
\mathcal{A}_{\text{FLT}}(\cos\theta_\qp) = \sum_l \bigl( A_l + B_l \:
\spin \bigr) \: P_l(\cos\theta_\qp) .
\end{equation}
The forward scattering amplitude is related to the Fermi liquid
parameters through~\cite{Landau3}
\begin{align}
A_l &= \frac{F_l}{1+F_l/(2l+1)} \label{af} \\[1mm]
B_l &= \frac{G_l}{1+G_l/(2l+1)} \label{bg} .
\end{align}
These relations are fulfilled by our numerical solution with an
accuracy of the order of $\Delta q/\kf$ and $\Delta q'/\kf$,
where $\Delta q$ and $\Delta q'$ are the momentum grid spacings in
$q$ and $q'$ respectively.
In Fig.~\ref{farel} we compare the numerical RG solution for the
forward scattering amplitude to that obtained from the corresponding
quasiparticle interaction using these relations,
Eqs.~(\ref{af},\ref{bg}). The antisymmetry of the
scattering amplitude is preserved with machine precision.

We start the discussion of the numerical results by presenting the
change of the effective mass, i.e., the ratio of the effective mass
obtained from the direct interaction to the full effective mass. The
ratio $m^\star(\la_0)/m^\star(\la)$ is taken as an approximation to
the running \zf~factor in the RG flow. The final value $m^\star(\la_0)
/m^\star(\la=0)$ is shown in Fig.~\ref{zfact}.
For Fermi momenta $1.0 \, \text{fm}^{-1} <
\kf < 1.5 \, \text{fm}^{-1}$, this approximation
agrees well with the values obtained in Brueckner-Hartree-Fock
calculations including rearrangement terms, as reported by Zuo {\em et
al.}~\cite{Zuo}. At lower densities, $0.5 \, \text{fm}^{-1} <
\kf < 1.0 \, \text{fm}^{-1}$, the agreement is qualitatively good, but
we find a somewhat smaller value of the \zf~factor compared with
the results of Baldo and Grasso~\cite{BG}. We notice that, for
densities below $\kf = 0.5 \, \text{fm}^{-1}$, this procedure yields
a quasiparticle strength slightly larger than one, while above $1.5
\,\text{fm}^{-1}$, the resulting \zf~increases with density.
The increase of \zf~with density cannot be excluded from first
principles, but seems unlikely. Thus, we conclude that our simple
procedure seems to work reasonably well in the density range $0.5
\,\text{fm}^{-1} <\kf < 1.5 \,\text{fm}^{-1}$. However, outside
this range, corrections due to e.g., rearrangement terms, are
probably needed. In this paper, our emphasis is on particle-hole
polarization effects.

\subsection{The Fermi liquid parameters and the scattering amplitude
at finite $q$}
\label{flparanm}

\begin{figure}[t]
\begin{center}
\includegraphics[scale=0.34,clip=]{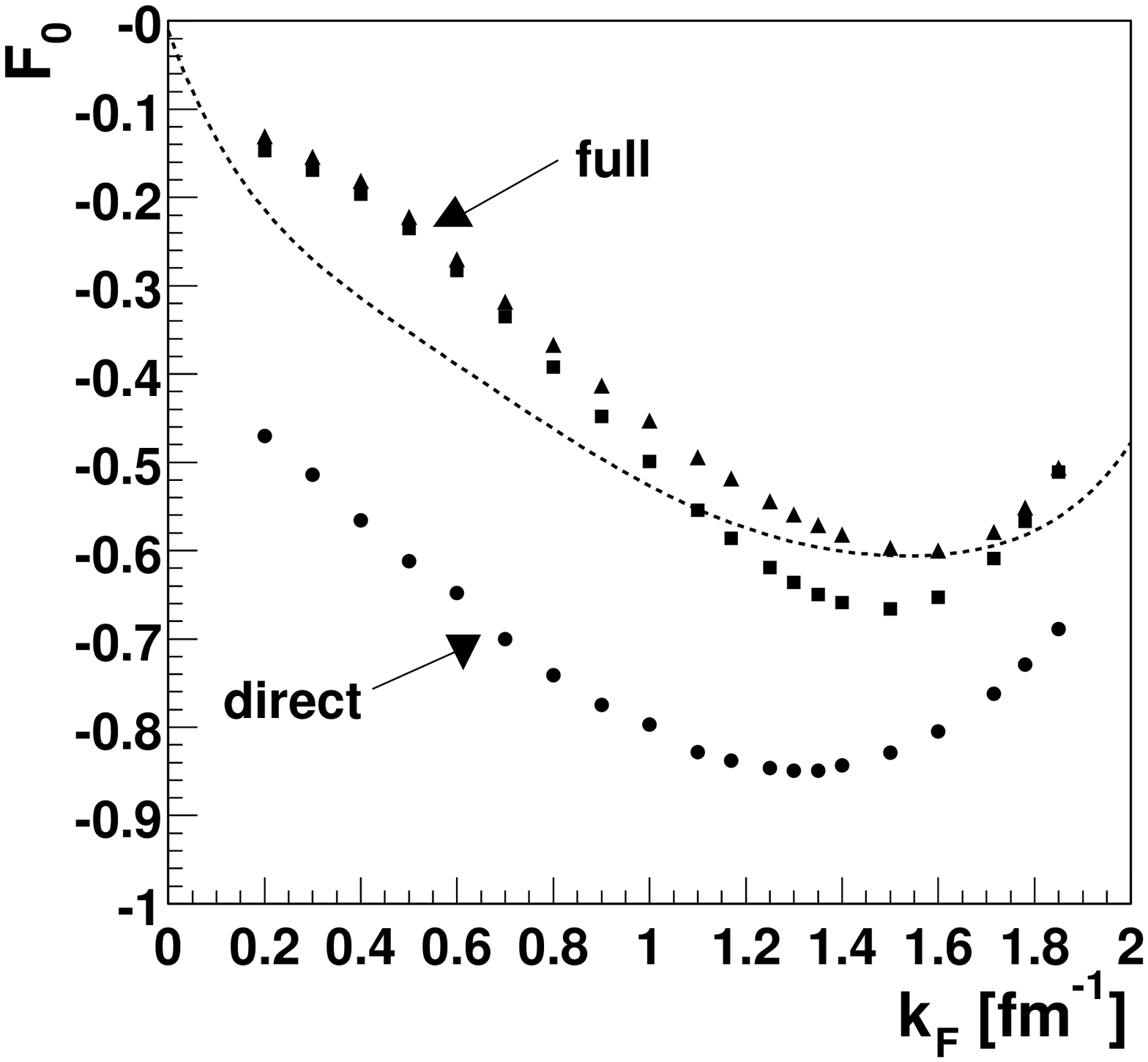}
\includegraphics[scale=0.34,clip=]{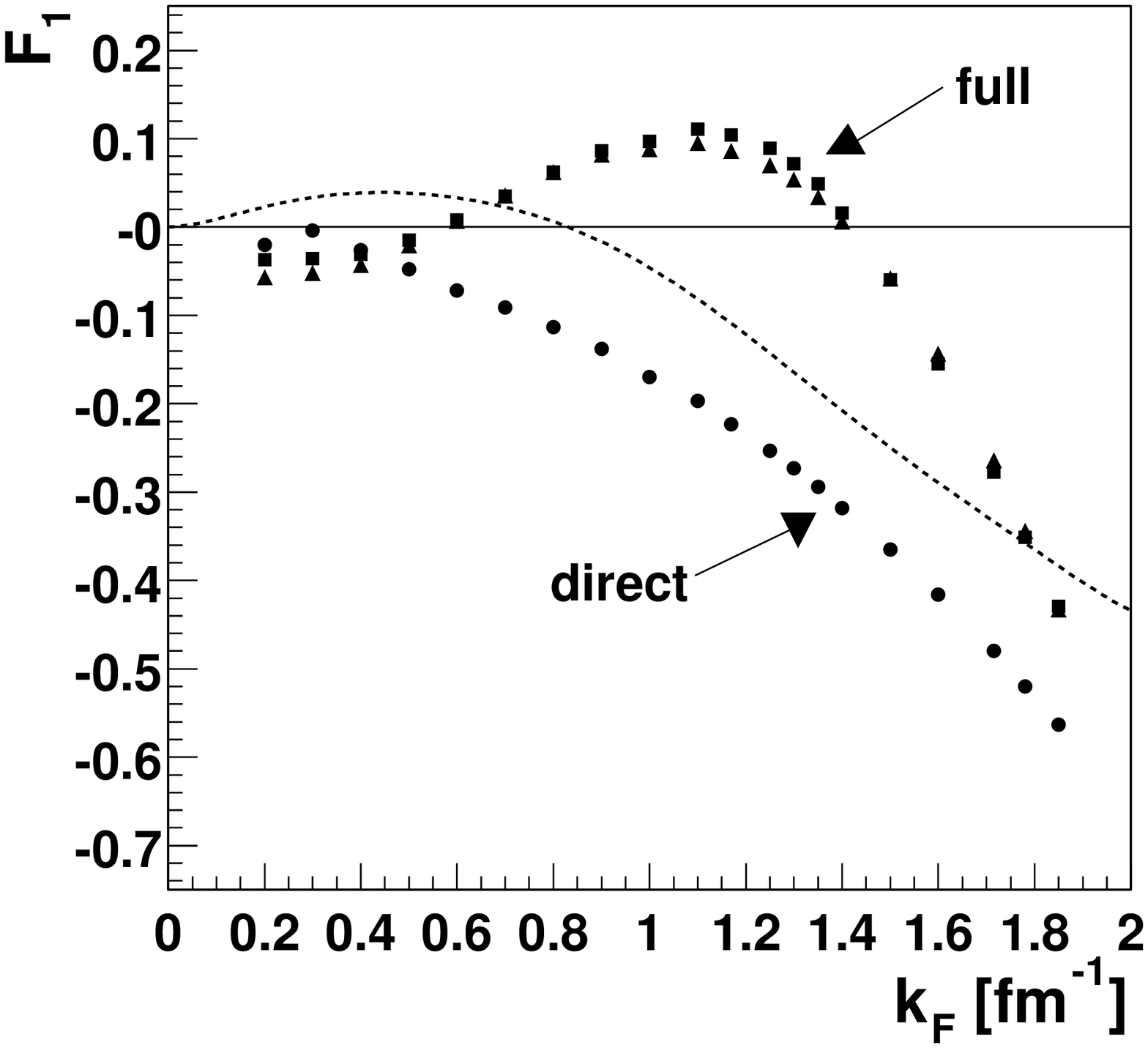}
\includegraphics[scale=0.34,clip=]{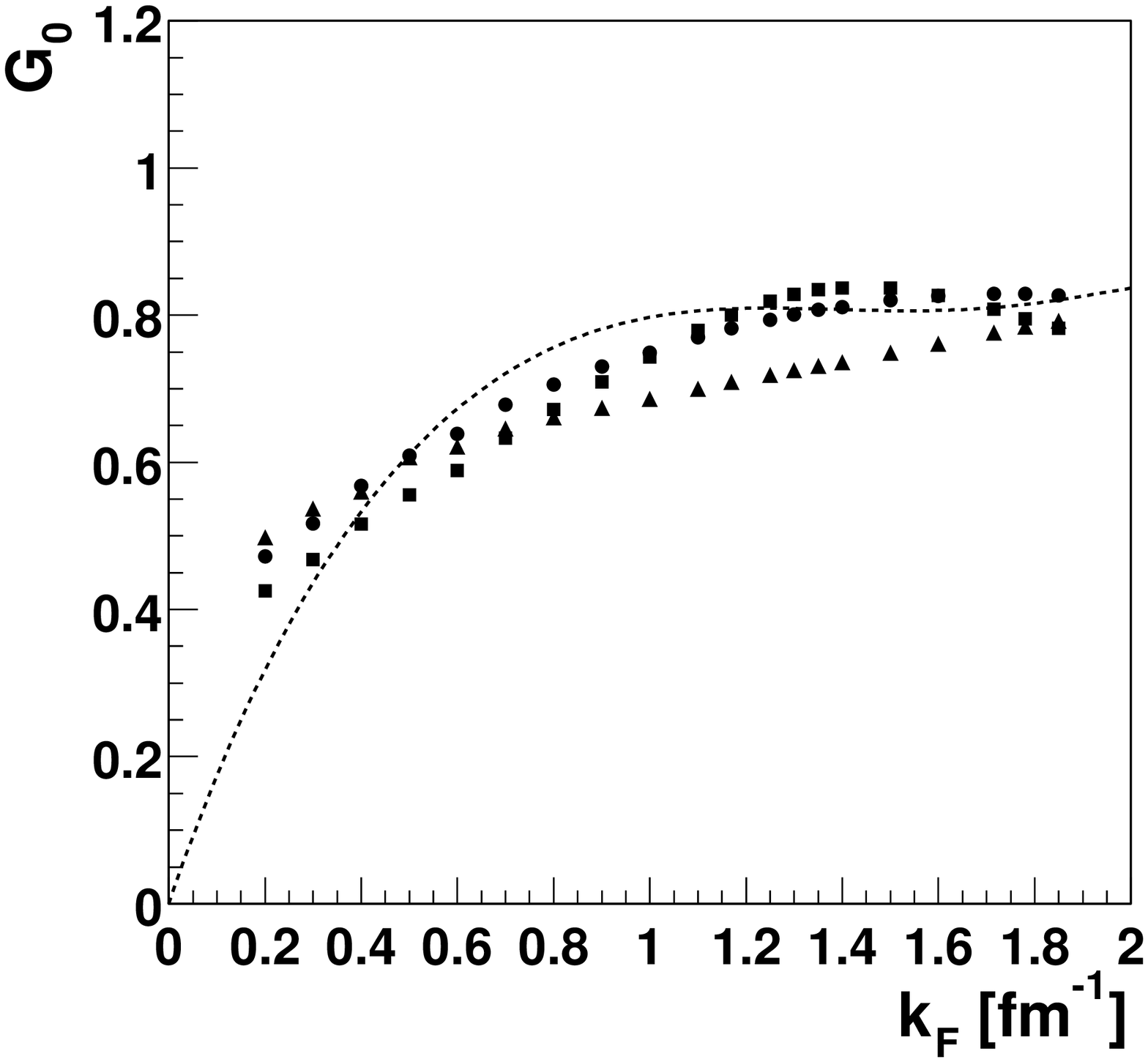}
\includegraphics[scale=0.34,clip=]{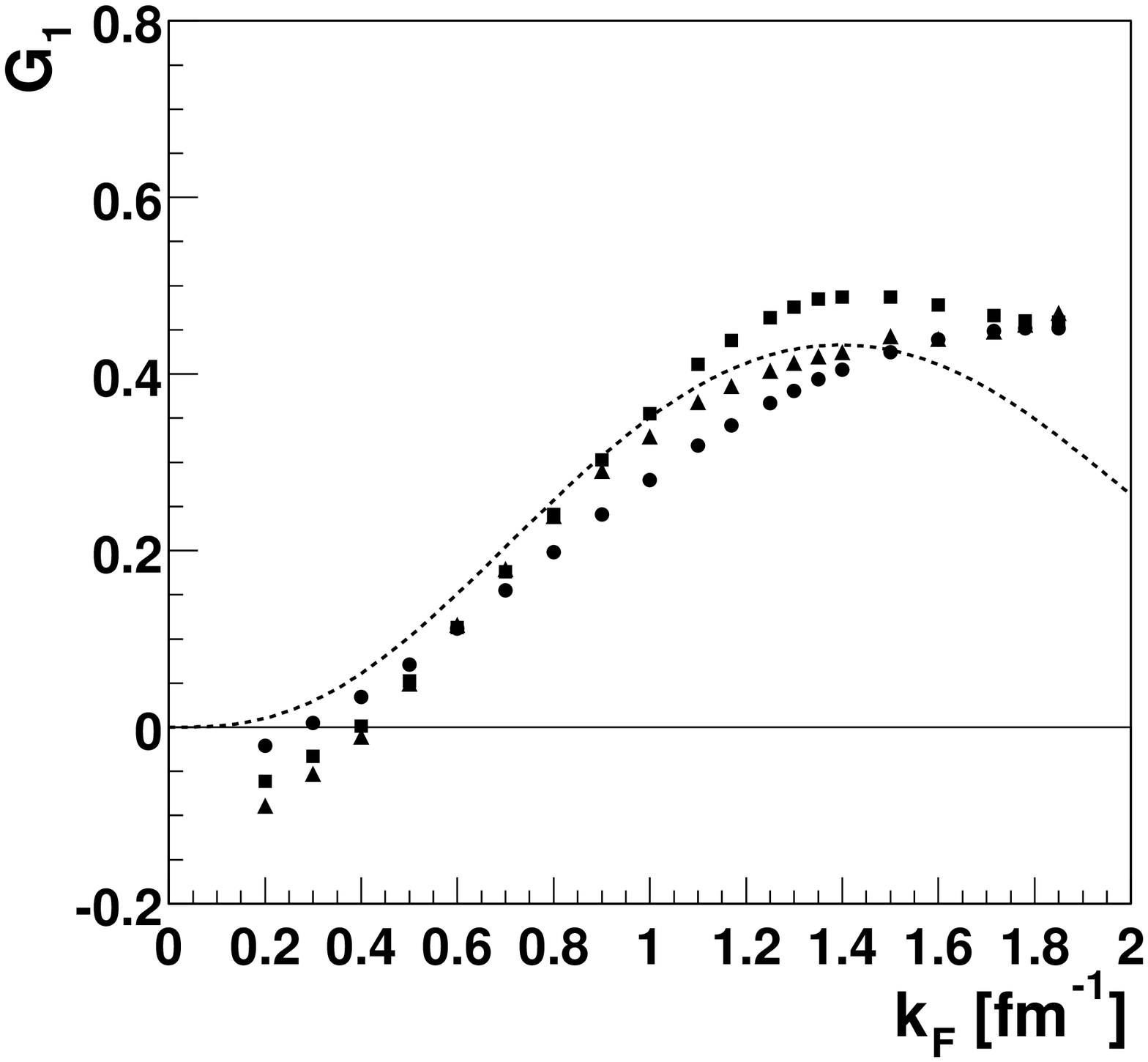}
\end{center}
\caption{The $l=0$ and $l=1$ Fermi liquid parameters versus the Fermi momentum
$\kf$. The dots denote the direct contribution only ($z_{\kf}=1$,
but including the effective mass in the density of states), whereas the
squares (constant \zf~factor) and the triangles (adaptive \zf~factor)
are calculated from the full RG solution. The results of Wambach {\em
et al.}~\cite{WAP2} are given for comparison as dashed lines.}
\label{Flp}
\end{figure}

In Fig.~\ref{Flp} we show our results for the $l=0$ and $l=1$ Fermi
liquid parameters as functions of the Fermi momentum. Since the RG
flow is antisymmetric by construction, the Fermi liquid parameters
obey the Pauli principle sum rules~\cite{Landau3,BD}. The low
momentum interaction drives the flow in the ZS' channel for the
quasiparticle interaction, and thus our results include effects of
the induced interaction of Babu and Brown.\footnote{By analyzing
the RG equations in~\cite{IIpaper}, one finds that the one-loop RG
generates a subset of the particle-hole parquet diagrams. The
diagrams that are missed correspond to terms, where the
renormalization is due to integrating out internal lines of the
vertex functions $\gamma$ in Fig.~\ref{zsandzsp}. In the parquet
equations, they correspond to certain diagrams involving left or
right vertex corrections in the direct or exchange particle-hole
channel.}
A further advantage of the RG approach is that the
calculations can be performed without truncating the expansion of
the quasiparticle interaction, Eq.~(\ref{legendre}), at some $l$.

Generally, we find a very good agreement between our
results\footnote{We note that at large Fermi momenta a
smaller \zf~factor, as discussed above, would lead to an overall
decrease of the magnitude of the Fermi liquid parameters.} and the
ones obtained using the polarization potential model by Ainsworth,
Wambach and Pines~\cite{WAP1,WAP2}. There are minor differences in
the value of the effective mass, which is treated self-consistently
in the RG approach. We note that, in the density range $0.6 \,
\text{fm}^{-1} < \kf < 1.4 \, \text{fm}^{-1}$, we find that the effective
mass at the Fermi surface exceeds unity. The quasiparticle
interaction was also calculated previously using the induced
interaction in~\cite{FLnm,JKMS,Baldo}. In these papers, the value
for the $l=0$ spin-dependent parameter $G_0$ is in good agreement
with ours, while for the spin-independent $F_0$, there are
differences. We stress the important role of the large $G_0$ for
the induced interaction. This Landau parameter causes the strong
spin-density correlations, which in turn enhance the Landau
parameter $F_0$ and consequently the incompressibility of neutron
matter. 

\begin{figure}[t]
\begin{center}
\includegraphics[scale=0.34,clip=]{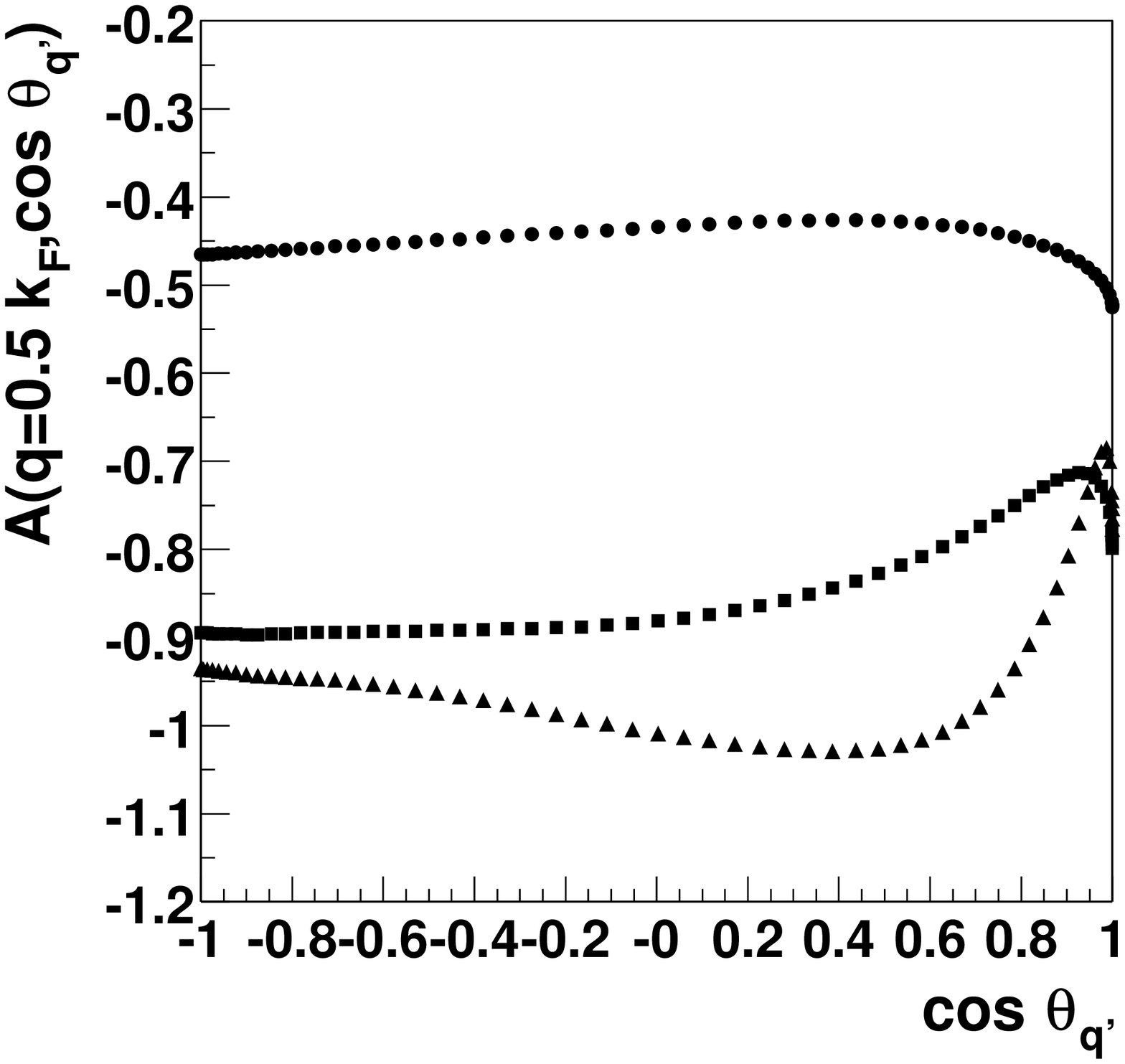}
\includegraphics[scale=0.34,clip=]{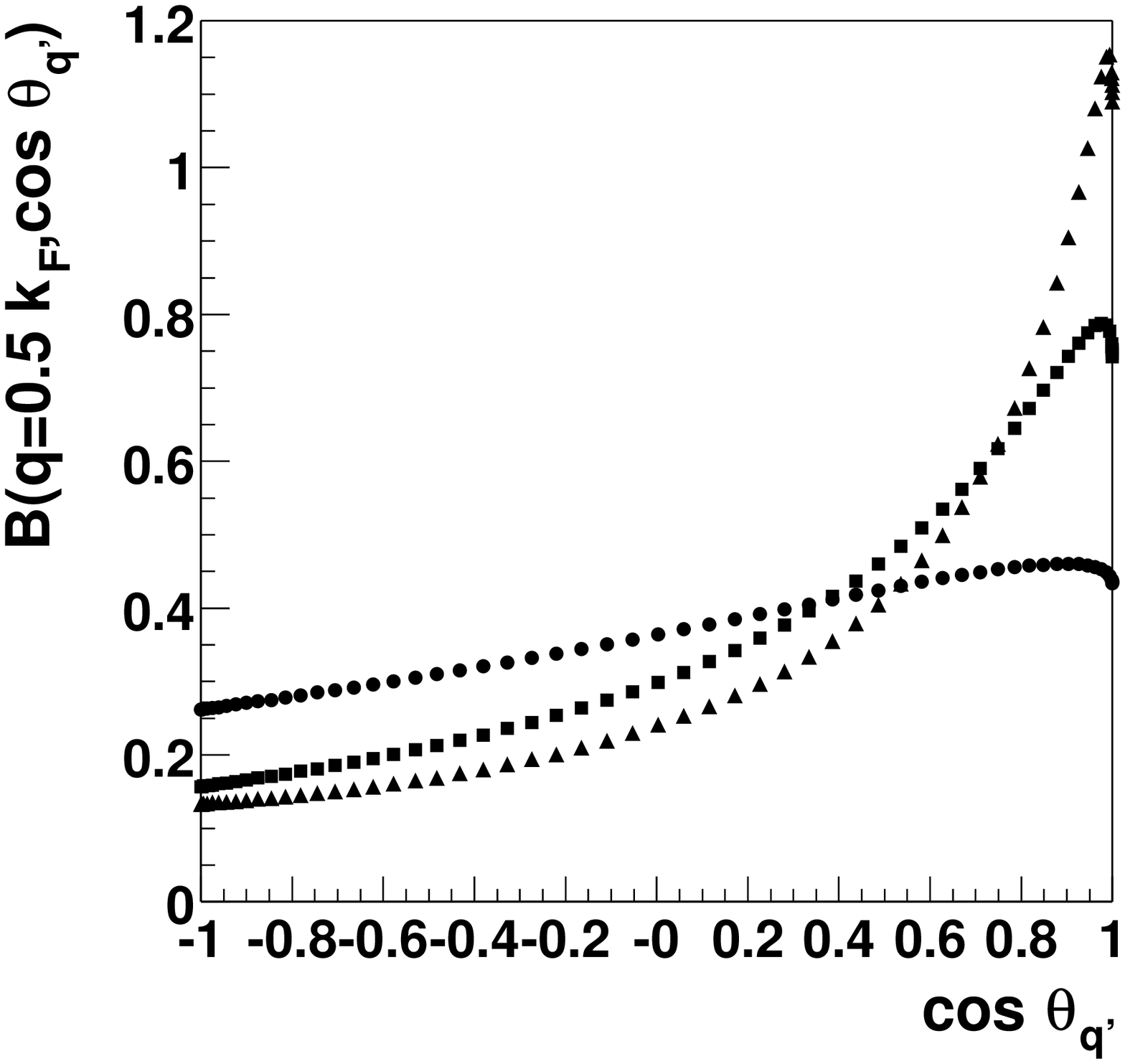}
\vspace*{-0.1mm}
\includegraphics[scale=0.34,clip=]{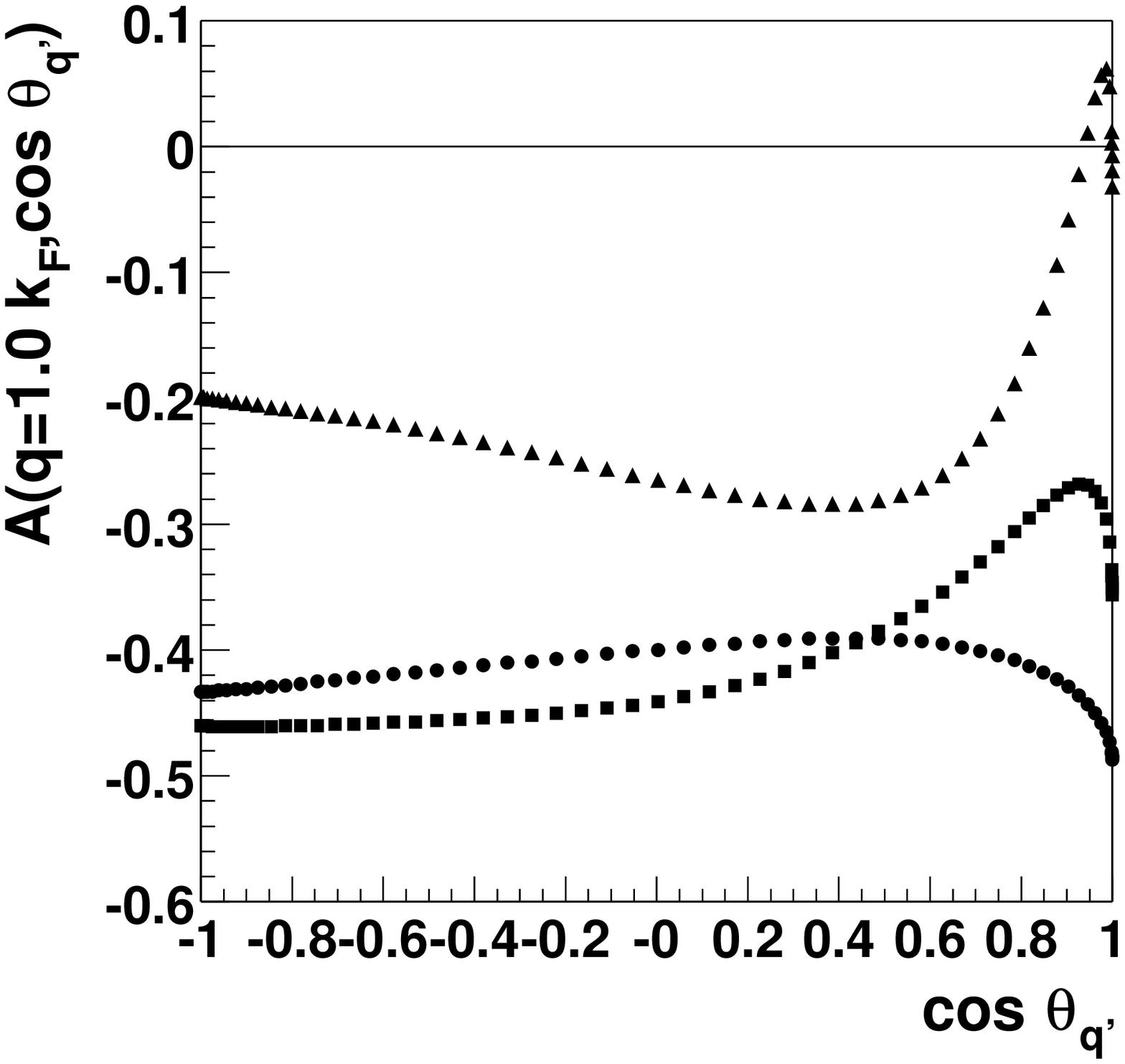}
\includegraphics[scale=0.34,clip=]{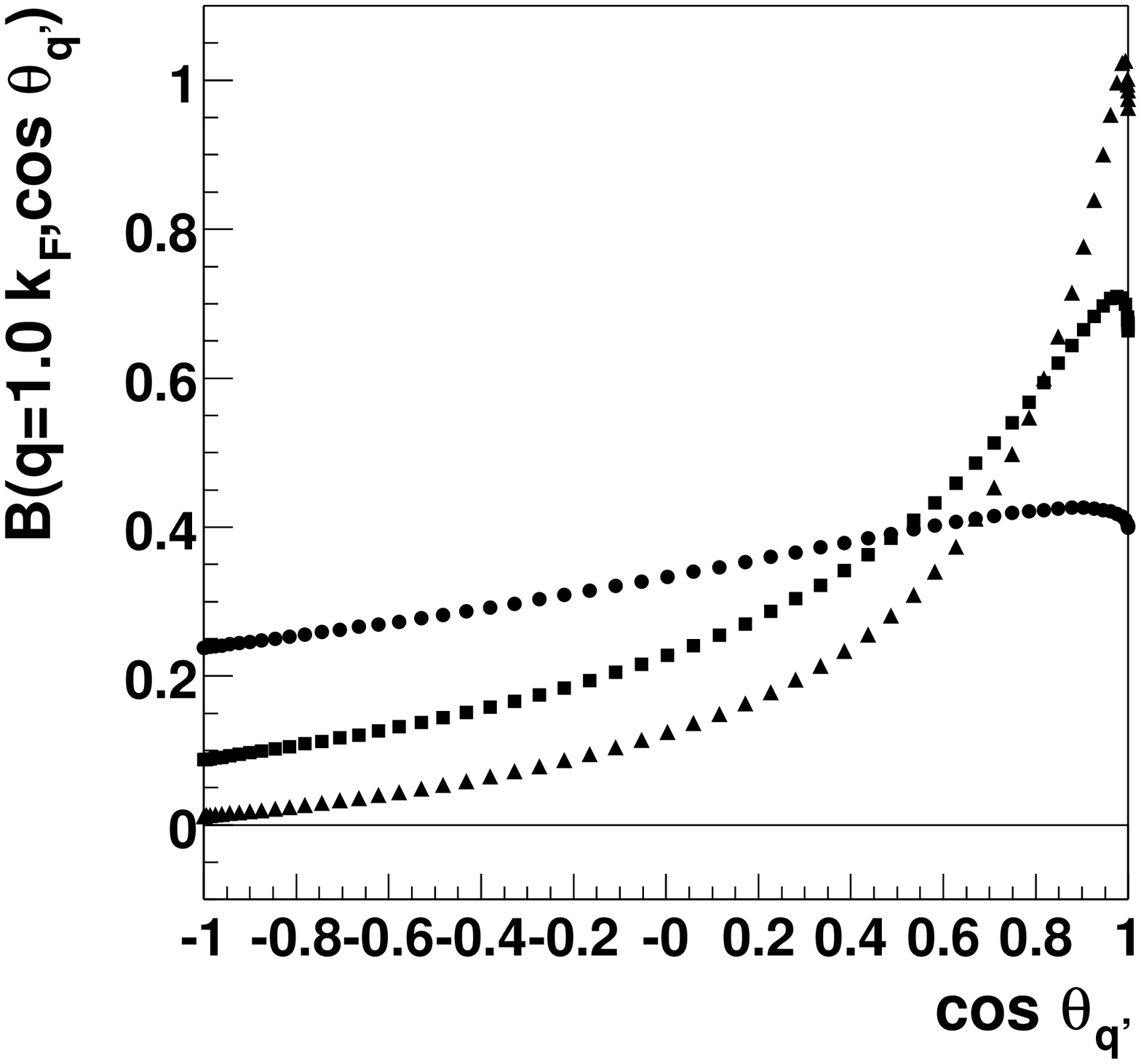}
\vspace*{-0.1mm}
\includegraphics[scale=0.34,clip=]{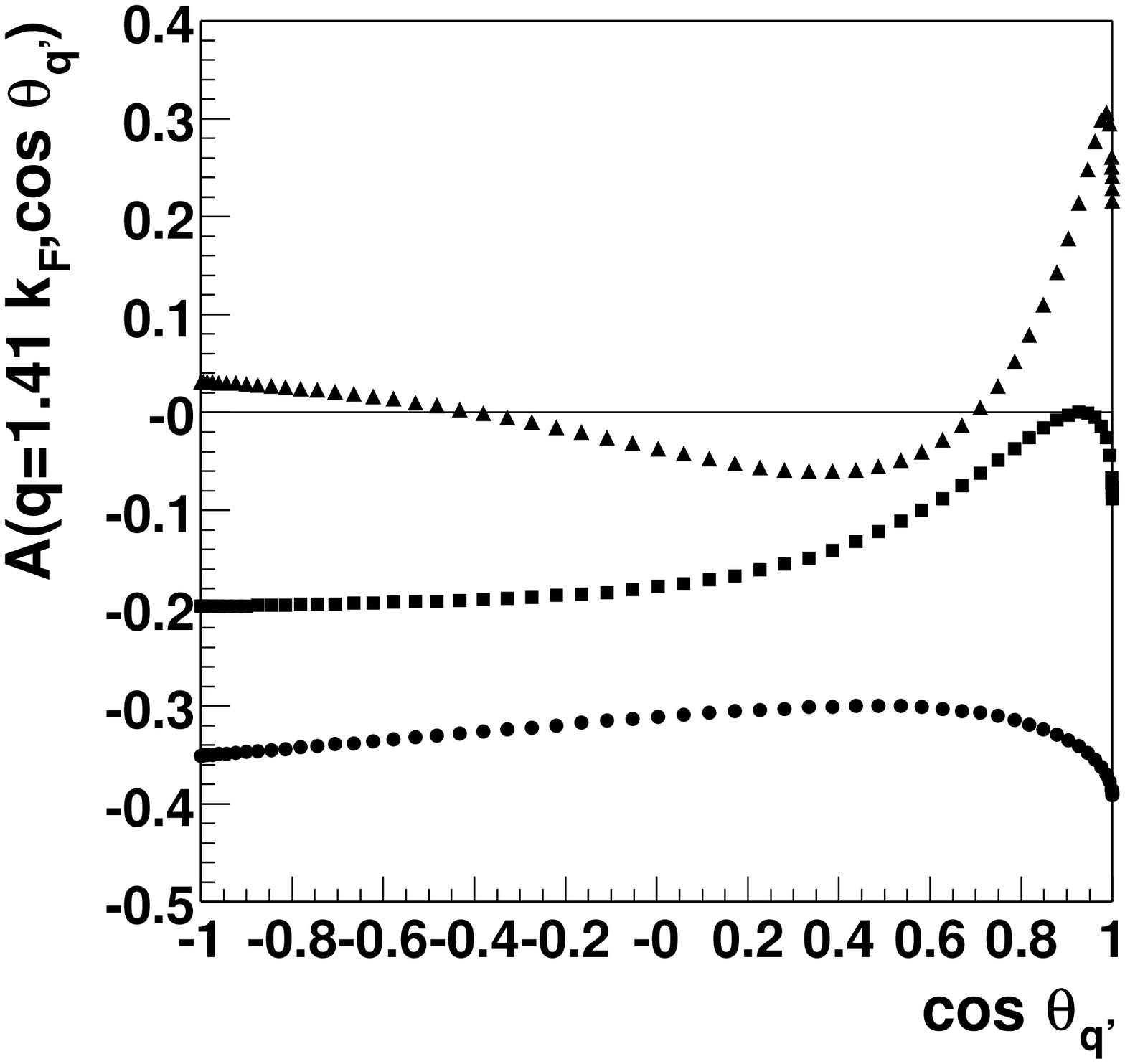}
\includegraphics[scale=0.34,clip=]{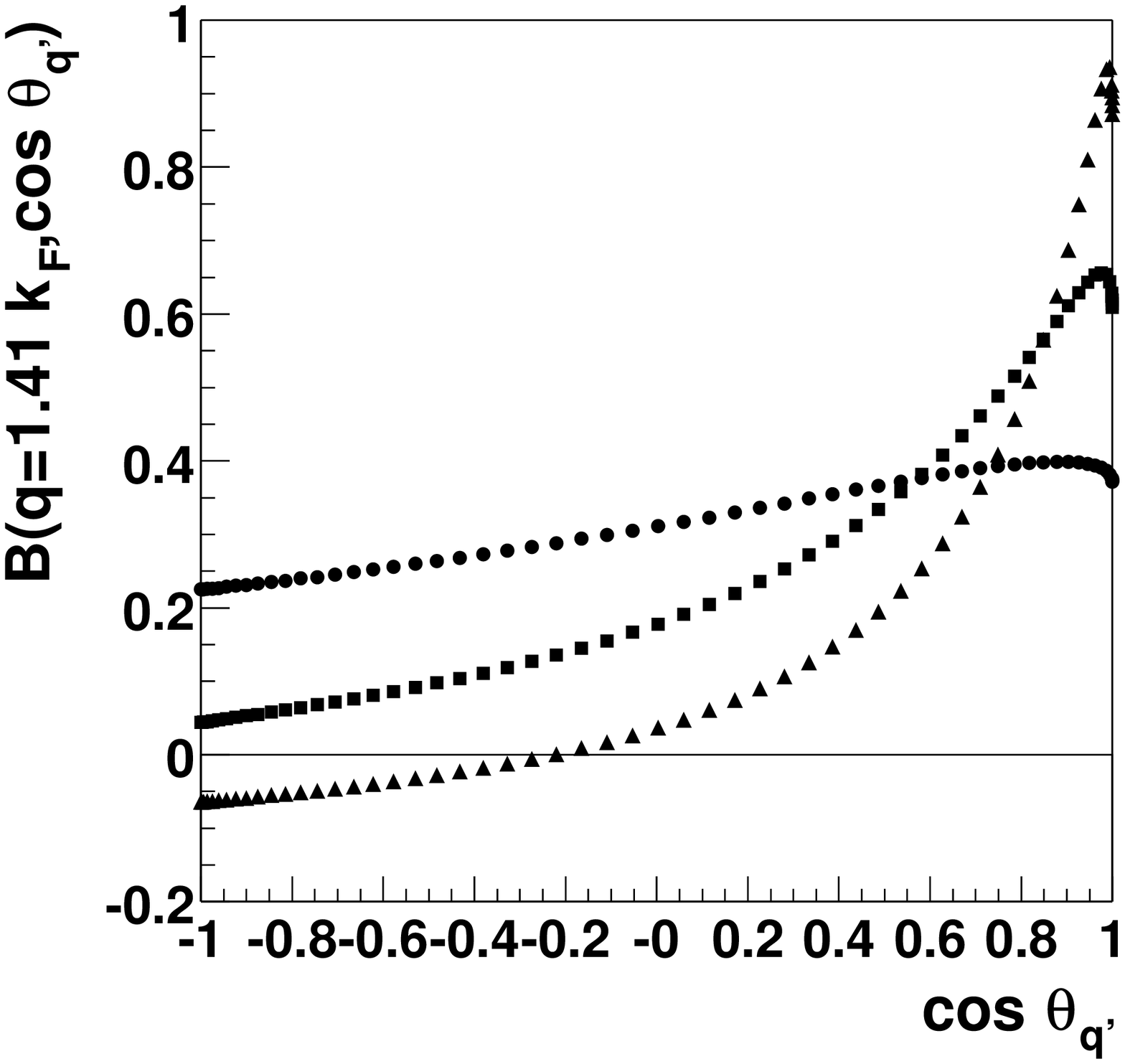}
\end{center}
\caption{The RG solution for the scattering amplitude on the
Fermi surface $A(\q,\qp)$ and $B(\q,\qp)$ versus $\cos\theta_\qp$ for
different cuts at finite momentum transfer $q = 0.5, 1.0$ and
$\sqrt{2} \, \kf$. The RG flow is solved with the adaptive \zf~factor
for different densities, $\kf = 0.6 \, \text{fm}^{-1}$ (dots), $\kf
= 1.0 \, \text{fm}^{-1}$ (squares) and $\kf = 1.35 \, \text{fm}^{-1}$
(triangles).}
\label{Asol}
\end{figure}
The results can be qualitatively understood by inspecting the
RG equation for the quasiparticle interaction, Eq.~(\ref{rgeqspin})
for $q=0$,
\begin{align}
\Lambda\,\frac{d}{d \Lambda} \, a(q=0,\qp;\la) &= - \Theta(q'-2\la) \: \bigl\{
\frac{1}{2} \, \beta_{\text{ZS'}}[a,\qp,\la] + \frac{3}{2} \,
\beta_{\text{ZS'}}[b,\qp,\la] \bigr\} \label{qpfrg} \\[2mm]
\Lambda\,\frac{d}{d \Lambda} \, b(q=0,\qp;\la) &= - \Theta(q'-2\la) \: \bigl\{
\frac{1}{2} \, \beta_{\text{ZS'}}[a,\qp,\la] - \frac{1}{2} \,
\beta_{\text{ZS'}}[b,\qp,\la] \bigr\} . \label{qpgrg}
\end{align}
In this qualitative argument we can neglect Fermi liquid parameters 
with $l \geqslant 1$. At a typical Fermi momentum $\kf =
1.0\,\text{fm}^{-1}$, we observe that the initial $F_0$ and $G_0$
are similar in absolute value, $|F_0| \approx |G_0| \approx 0.8$.
Consequently, there is a cancellation between the
contributions due to the spin-independent and spin-dependent
parameters in Eq.~(\ref{qpgrg}), while in Eq.~(\ref{qpfrg}) both
contributions are repulsive. Thus, one expects a relatively small 
effect of the RG flow on $G_0$ and a substantial renormalization of 
$F_0$, in agreement with our results.

The approximate relationship for the \zf~factor,
Eq.~(\ref{zfactrel}), is reasonable, if there is a strong
frequency dependence of the polarization contributions to the
nucleon self-energy. This may be the case if one of the Fermi
liquid parameters is strongly attractive~\cite{Kevin}. In the
density range where we obtain a reasonable \zf~factor, $F_0$ is
indeed most negative.

The RG approach enables us to compute the scattering amplitude for
general scattering processes on the Fermi surface, without making
further assumptions for the dependence upon the particle-hole
momentum transfers $q$ and $q'$. In Fig.~\ref{Asol}, we show the
full scattering amplitude for different values of the momentum
transfer $q$ and different densities. We recall that for scattering
on the Fermi surface the exchange momenta are restricted to
$q^2+q^{\prime 2} \leqslant 4 \, \kf^2$. The scattering amplitude at
finite momentum transfer is of great interest for calculating
transport processes and for assessing the consequences of
superfluidity in neutron stars.

\subsection{The $^1$S$_0$ superfluid gap}
\label{gapsec}

It has been discussed by several authors that polarization effects,
which are necessary in order to satisfy the Pauli principle in
microscopic calculations, have a profound influence on
superfluidity in neutron stars~\cite{WAP1,WAP2,Clark,3pf2}. A
reduction of the pairing gap due to polarization effects was first
found by Clark {\em et al.}~\cite{Clark}. Subsequently, Ainsworth,
Wambach and Pines explored particle-hole polarization effects in
detail, using the framework of the induced interaction~\cite{BB}
combined with the polarization potential model~\cite{WAP1,WAP2}.
They found a large reduction of the $^1$S$_0$ superfluid gap from a
maximum gap of $5 \, \text{MeV}$ to $0.9 \, \text{MeV}$, when the
induced interaction was included. A reduction to similar magnitude
was obtained by Chen {\em et al.} in a variational calculation,
when CBF correlations were included~\cite{Chen1,Chen2}.
Recently, Schulze {\em et al.} also found a
substantial reduction of the gap by employing a phase-space
averaging method to extrapolate the scattering amplitude to
back-to-back kinematics~\cite{Baldo}.

Nevertheless, when one compares these calculations in more detail,
one finds that the maximum gap and the density dependence vary
appreciably among the different approaches (for a review on
superfluidity of neutron matter and a compilation of the
results, see the recent notes by Lombardo and
Schulze~\cite{LomSch}). Consequently, further studies are needed in
order to improve the microscopic calculations and in order to make reliable
predictions of superfluid properties of neutron stars. The RG
approach offers a systematic framework for computing the
quasiparticle scattering amplitude, which is needed in the
calculation of the superfluid gap. In this paper we have presented
a straightforward application of the RG method, where both momentum
scales, $q$ and $q'$, are treated on the same footing, and
consequently the antisymmetry of the scattering amplitude is
preserved. Moreover, the RG method can provide a unifying and
transparent framework for the particle-particle and particle-hole
effects. Finally, by using the unique low momentum interaction
$V_{\text{low k}}$ as the direct interaction, much of the model
dependence is removed.

We estimate the $^1$S$_0$ pairing gap using weak coupling BCS
theory. The pairing matrix element averaged over the Fermi surface
is given by\footnote{The factor $1/8$ in Eq.~(\ref{pairel})
consists of a factor $1/2$ for the average over $\cos\theta_\q$,
another factor $1/2$ for the reduction of the full density of
states to that of one spin species. The remaining factor $1/2$
accounts for the normalization of the antisymmetrized contact
interaction employed in the standard derivation of the weak
coupling result (see e.g.,~\cite{AGD}).}
\begin{equation}
\langle \mathcal{A} \rangle = \frac{1}{8} \int_{-1}^{1} d
\cos\theta_\q \: \bigl\{ A(\,q,\sqrt{4 \kf^2 - q^2}\,) - 3 \: B
(\,q,\sqrt{4 \kf^2 - q^2}\,) \bigr\} ,
\label{pairel}
\end{equation}
where the momenta for back-to-back scattering are constrained by
$q^2 + q^{\prime 2} = 4 \, \kf^2$. An estimate of the superfluid
gap is then given by the weak coupling formula~\cite{PZ}
\begin{equation}
\Delta = 2 \: \varepsilon_{\text{F}} \: \exp\biggl( \frac{1}{\langle
\mathcal{A} \rangle} \biggr) ,
\label{weakgap}
\end{equation}
where $\varepsilon_{\text{F}} = \kf^2 / 2 m^\star$ denotes the
Fermi energy and we use the same prefactor as Ainsworth {\em et
al.}~\cite{WAP1,WAP2}. For liquid $^3$He, the weak coupling BCS
results for the critical temperature are in remarkable agreement
with experiment when experimental Fermi liquid parameters are used
as input~\cite{PZ,BedPin}. For the direct interaction, the pairing
gap, Eq.~(\ref{weakgap}), with a free single particle spectrum is
given by
\begin{equation}
\Delta = \frac{\kf^2}{m} \: \exp\biggl( \cfrac{\pi}{2 \, \kf \: m \:
V_{\text{low k}}(\kf,\kf;\la=\sqrt{2} \: \kf; \text{$^1$S}_0) }
\biggr) .
\label{directgap}
\end{equation}
\begin{figure}[t]
\begin{center}
\includegraphics[scale=0.35,clip=]{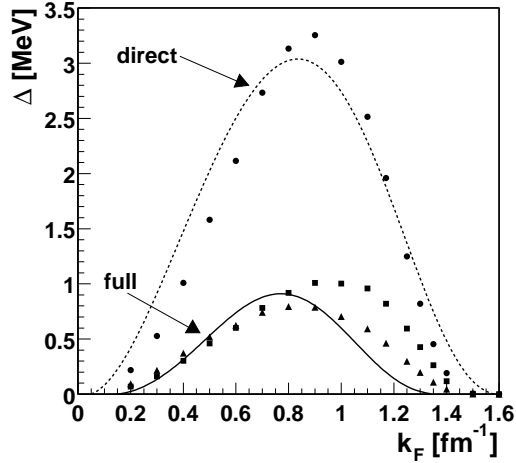}
\end{center}
\caption{The $^1$S$_0$ superfluid gap versus the Fermi momentum
$\kf$. The dots denote the direct contribution only ($m^\star/m
= z_{\kf} = 1$), whereas the squares (constant \zf~factor) and the
triangles (adaptive \zf~factor) are calculated from the full RG
solution. In comparison we show the values of the gap calculated by
solving the full BCS equation (dashed line), taken from~\cite{LomSch}. The
superfluid gap including particle-hole polarization effects
obtained by Wambach {\em et al.}~\cite{WAP2} is shown as a solid
line.}
\label{1s0gaps}
\end{figure}

In Fig.~\ref{1s0gaps} we show the superfluid gap computed with the
direct term only, Eq.~(\ref{direct}), as well as that obtained with
the RG solution for the scattering amplitude. To facilitate the
comparison with other calculations, the gap obtained with the
direct interaction is given for a free single particle spectrum,
i.e., $m^\star/m = z_{\kf} = 1$. The full solution of the BCS
equation using various free nucleon-nucleon interactions yields a
maximum gap of about $3\,\text{MeV}$ at a Fermi momentum of $0.8-
0.9\, \text{fm}^{-1}$ (for an overview of the results see Fig.~2
in~\cite{LomSch}; see also~\cite{Cugnon,Khodel,Elga}). The
agreement of the weak coupling result using $V_{\text{low k}}$,
Eq.~(\ref{weakgap}), with these calculations is remarkably good.
The reason for this success is presumably that in 
$V_{\text{low k}}$, the strong short-range repulsive 
core is eliminated, whereas the phase shifts are preserved.
For such a ``weak'' interaction, the weak coupling treatment is
expected to work.

In the full RG calculation, including particle-hole polarization
effects, we find a much smaller gap; the maximum gap is $0.8 \,
\text{MeV}$ at a Fermi momentum of $\kf = 0.8 \, \text{fm}^{-1}$.
Here, both the effective mass and the quasiparticle strength
$z_{\kf}^2$ are taken into account in the quasiparticle scattering
amplitude. The gap is in good
agreement with that obtained by Wambach {\em et al.}~\cite{WAP2},
which is shown for comparison in Fig.~\ref{1s0gaps}. The results
obtained in other calculations, which include polarization effects,
are compiled in Fig.~7 of~\cite{LomSch}.

We observe that the full gap vanishes at $\kf = 1.45 \, \text{fm}^{-1}$
both for a constant and an adaptive \zf~factor. Moreover, at the
lowest Fermi momentum we considered, $\kf = 0.2\,\text{fm}^{-1}$, the
ratio of the full to the direct gap is given by $\Delta/\Delta_0 =
0.43$, which is close to the universal low density limit $\Delta/\Delta_0 = (4
e)^{-1/3}\approx 0.45$~\cite{Gorkov}.\footnote{We note that the low
density limit for the ratio of the full to the direct S-wave gap is
independent of the scattering length.
While the strict low density expansion in $\kf a_{\text{S}}$ is not
relevant to neutron star matter, since its validity is restricted
to densities below neutron drip, an expansion in an effective
scattering length $\kf a_{\text{eff}}$ may be valid up to much
higher densities. As argued in~\cite{IIpaper}, $V_{\text{low
k}}(0,0;\la_{V_{\text{low k}}}
\sim \kf)$ can be interpreted as the effective scattering length in
the particle-particle channel. In the isotriplet channel
$V_{\text{low k}}(0,0) \approx 2.0 \, \text{fm}$ in the cutoff
independent region. If one adopts the effective scattering length,
it is not surprising to find a ratio close to the universal one at
$\kf=0.2\,\mbox{fm}^{-1}$, since $\kf \ll 1/|a_{\text{eff}}|\approx
0.5\,\text{fm}^{-1}$.}

In the direct interaction, $V_{\text{low k}}$, particle-particle ladder
diagrams are included for large relative momenta, $k >
\la_{V_{\text{low k}}}$. In fact, we argued that the cutoff
accounts for the Pauli blocking of intermediate particle states.
This raises the question of double counting, since the BCS equation
also sums particle-particle ladders. However, since the gap is
generated by low-lying two-particle and two-hole states, which are
excluded in $V_{\text{low k}}$ due to the cutoff at $\sqrt{2} \:
\kf$, the double counting problem is minimal. Furthermore, because
the cutoff dependence of $V_{\text{low k}}$ is very weak in the
density range relevant for superfluidity, any residual double
counting, due e.g., to states above the cutoff that contribute to
the solution of the BCS gap equation, would not affect our results
significantly.

We note that by using a density dependent cutoff, we recover the
correct low density limit, where the effective neutron-neutron
interaction is given by the free $T$ matrix. This follows because
the low momentum interaction $V_{\text{low k}}$ approaches the $T$
matrix for momenta below the inverse S-wave scattering length,
i.e., $\la_{V_{\text{low k}}}\sim \kf \ll 1/|a_{\text{S}}|$, as
shown in Fig.~5 of~\cite{Vlowkflow}. In particular, $V_{\text{low
k}}(\kf,\kf) \to a_{\text{S}}/m$, as $\kf \to
0$,\footnote{In~\cite{Vlowkflow} conventional units are used where
$\hbar^2/m = 1$.} and we recover the well known low density results
in Eq.~(\ref{directgap}), where the gap is given by the phase
shifts~\cite{Emery}. However, due to the large neutron-neutron
scattering length, the low density expansion is applicable only for
Fermi momenta below $\kf \ll 1/|a_{\text{S}}|=
0.05\,\text{fm}^{-1}$, which corresponds to extremely small
densities, well below neutron drip. Since neutron star matter at
such densities is in a solid phase, the strict low density
expansion in $\kf a_{\text{S}}$ is not relevant for the physics of
neutron stars.

Finally, we remark that the physical pairing gap is given by $z_{\kf} \,
\Delta$. Thus, our results should be multiplied by the quasiparticle
strength given in Fig.~\ref{zfact}.\footnote{Given the uncertainty
in the pre-exponential factor of Eq.~(\ref{weakgap}), this may seem
superfluous. However, for a given prefactor, it leads to an
additional suppression of the gap.}
We stress that the physical effect responsible for the strong
reduction of the gap is the polarization by the many-body medium. A
detailed analysis of the influence of self-energy contributions on
the superfluid gap can be found in~\cite{BG,Bozek,Schuck}.

Particle-hole polarization may have a strong effect also on the
color superconducting gap in high density QCD. We plan to explore
these effects using the methods presented here in future studies.

\subsection{The equation of state of neutron matter}

In contrast to nuclear matter, neutron matter does not saturate. It
is stable only if the positive pressure is balanced by an external
force, as e.g., gravitation in a neutron star. Therefore, it is
possible to compute the equation of state of neutron matter using
the Fermi liquid parameters presented in Section~\ref{flparanm}.
This method, which accounts for polarization effects on the equation of state,
was first employed by K\"allman~\cite{Kaellman}.

The speed of sound $c_{\text{S}}$ of a Fermi liquid is given by
the incompressibility $K$~\cite{Nozieres}
\begin{equation}
c_{\text{S}}^2 = \frac{d P}{d \rho} = \frac{1}{9 \, m} \: K ,
\end{equation}
where $\rho$ denotes the mass density and $K$ is the incompressibility. The
latter is in Fermi liquid theory given by
\begin{equation}
K = \frac{3 \, k_{\text{F}}^2 \: ( 1 + F_0 )}{m \: ( 1 + F_1 /
3 )} .
\end{equation}
In the density range of interest, the neutrons may to a good
approximation be treated non-relativistically. Thus, the mass
density is given by $\rho= m \, n$, where $n=\kf^3/3 \pi^2$ is the
number density. Now, we can solve the differential equation
\begin{equation}
\frac{d P}{d n} = \frac{1}{9} \, K
\label{diff1}
\end{equation}
for the equation of state $P(n)$. The incompressibility is computed at
each density using the Fermi liquid parameters obtained in the RG
approach. The energy per neutron is then computed by integrating
the thermodynamic relation
\begin{equation}
P = n^2 \, \frac{d (E/A)}{d n} .
\label{diff2}
\end{equation}

This approach requires two initial conditions for the differential
equations, Eqs.~(\ref{diff1},\ref{diff2}). One can match our results
with a conventional equation of state $P_{\text{C}}(n)$ at a certain
matching density $n_{\text{M}}$. For $n > n_{\text{M}}$, the
resulting equation of state $\widetilde{P}(n)$ reads
\begin{equation}
\widetilde{P}(n) = P(n) - P(n_{\text{M}}) + P_{\text{C}}(n_{\text{M}}) ,
\end{equation}
with a corresponding equation for the energy per particle. We match
the pressure and the energy to the $V_{\text{low k}}$
Hartree-Fock equation of state
at the density $n = 0.02 \, \text{fm}^{-3}$. This is the lowest
density given by Akmal {\em et al.}~\cite{APR}.\footnote{At very
low densities the Hartree-Fock calculation cannot be trusted due to
the large neutron-neutron scattering length, as discussed above.}
We stress that the equation of state depends on the matching
condition. The energy is particularly sensitive, since it requires
two constants of integration. Below we give a parametrization of
the pressure, with the boundary condition $P(n=0)=0$,
\begin{multline}
P(n) = 79.09 \: \biggl(\frac{n}{\text{fm}^{-3}}\biggr)^{5/3} \biggl\{ \:
1.000 - 1.177 \: \biggl(\frac{n}{\text{fm}^{-3}}\biggr)^{1/3}  \\[2mm]
- 0.324 \: \biggl(\frac{n}{\text{fm}^{-3}}\biggr) + 5.290 \:
\biggl(\frac{n}{\text{fm}^{-3}}\biggr)^2 \: \biggr\} \:
\frac{\text{MeV}}{\text{fm}^{3}} \, .
\end{multline}
With this parametrization, the reader can match our equation of
state to a conventional one at low densities.

\begin{figure}[t]
\begin{center}
\includegraphics[scale=0.34,clip=]{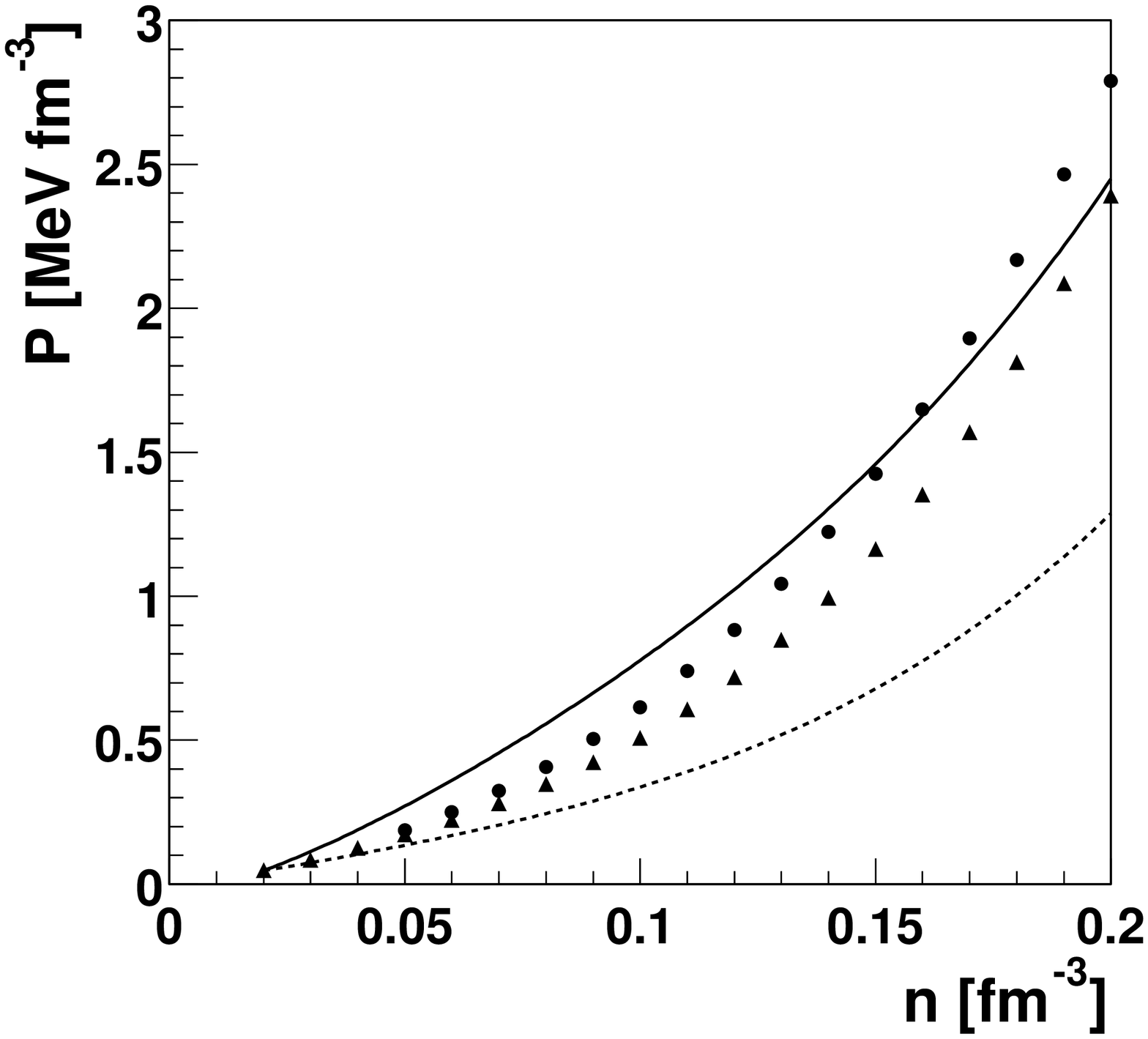}
\includegraphics[scale=0.34,clip=]{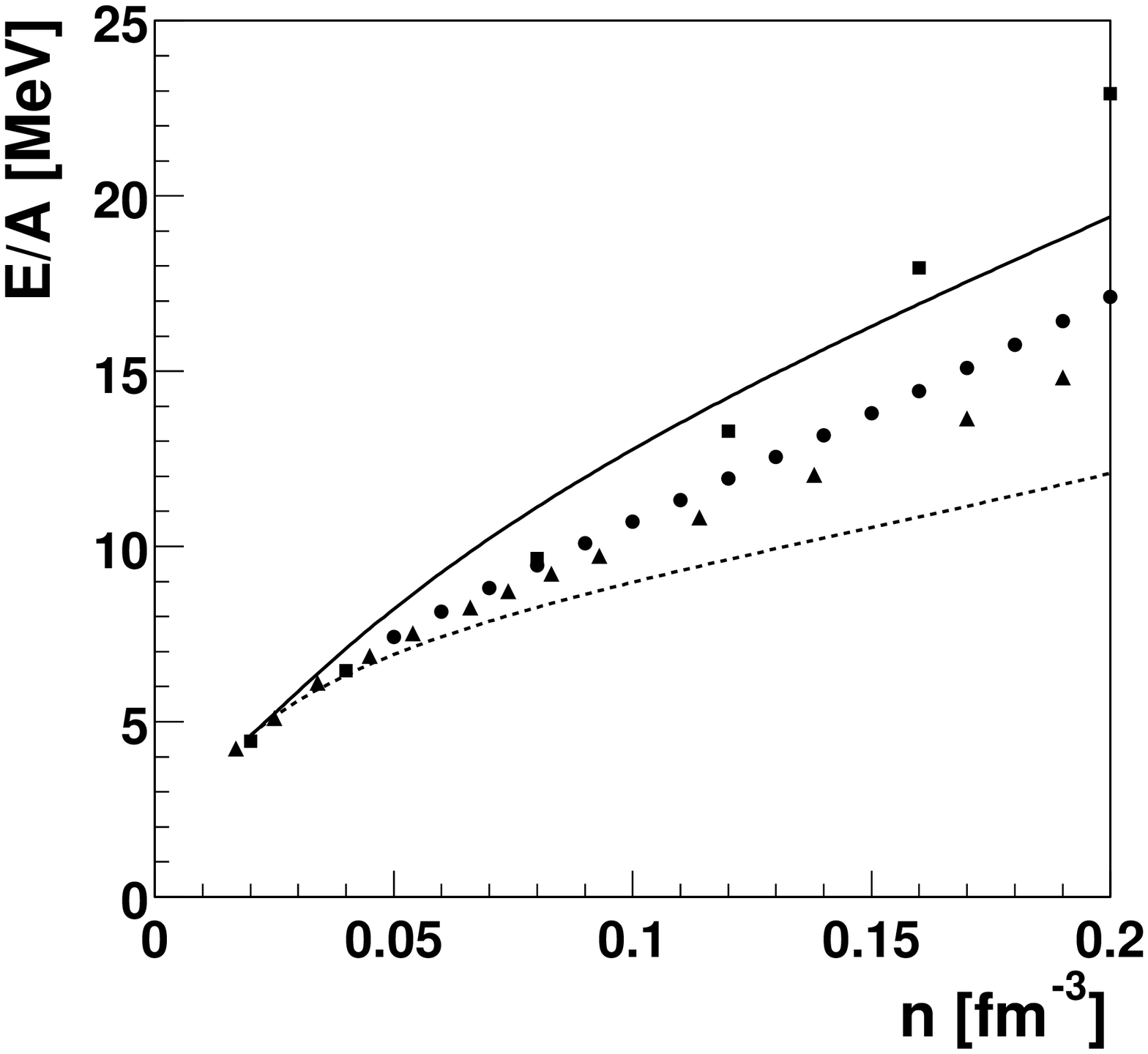}
\end{center}
\caption{The pressure $P$ and the energy per neutron $E/A$ for pure
neutron matter as functions of density $n$. Results are given for
the full Fermi liquid parameters (solid line) and the direct ones
(dashed line) matched to the $V_{\text{low k}}$ Hartree-Fock
equation of state at low density, $n=0.02 \, \text{fm}^{-1}$.
Also given is the $V_{\text{low k}}$ Hartree-Fock contribution to the
energy~\cite{thesis} and the pressure calculated by means of
Eq.~(\ref{diff2}) (triangles). In comparison, we show
the Brueckner-Hartree-Fock equation of state of Bao {\em et
al.}~\cite{Engvik} (dots) computed with the non-relativistic
Bonn potential and the equation of state of Akmal {\em et
al.}~\cite{APR} (squares) from a variational calculation using the
Argonne $v_{18}$ interaction supplemented by three-body forces and
the boost corrections. For a comparison with other equations of
state, see also Fig.~1 in Lattimer and Prakash~\cite{LP}.}
\label{eos}
\end{figure}
The resulting pressure and the energy per neutron are shown in
Fig.~\ref{eos}. We note that the Hartree-Fock energy obtained with
$V_{\text{low k}}$ and the energy obtained from the direct Landau
parameters do not coincide. This is solely due to the density dependence
of the cutoff $\la_{V_{\text{low k}}}$, which gives rise to additional
contributions to the effective interaction. However, since a major
part of these contributions correspond to particle-hole propagation
in the ZS' channel, they are to a large extent part of the induced
interaction and should not be included in the direct term.
Consequently, the difference between the Hartree-Fock energy and that
obtained using the full induced interaction is due to higher order
particle-hole polarization effects~\cite{BBfornuclmat}.

Relative to the $V_{\text{low k}}$ Hartree-Fock
calculation~\cite{thesis} there is an increase in pressure and
energy below nuclear matter density. Compared to other
non-relativistic equations of state~\cite{Engvik,APR}, the pressure
and energy is somewhat larger at low densities.\footnote{The equation
of state of Akmal {\em et al.} is significantly stiffer at higher
densities due to three-body forces.} At densities above
$0.11 \, \text{fm}^{-3}$, corresponding to Fermi momenta $\kf > 1.5 \,
\text{fm}^{-1}$, the resulting \zf~factor is larger than that
found in Brueckner-Hartree-Fock calculations, as noted above. A
smaller \zf~factor would reduce the attractive contribution from
$F_0$ and thus lead to an increase of the incompressibility and
consequently also of the pressure and the energy for $n >
0.11\,\text{fm}^{-3}$ in our calculation. We conclude that
polarization effects possibly have an important impact on the
equation of state, which should be included in a consistent treatment
of neutron star superfluidity and neutron star structure.

\section{Summary and conclusions}

In this work, we have developed a practicable framework for
renormalization group calculations of strongly interacting Fermi
systems. We employed the RG approach to
compute the quasiparticle interaction and the scattering amplitude on
the Fermi surface for neutron matter, which for many purposes is a
good approximation to neutron star matter, where there is a small
admixture of protons. Our approach follows the ideas of Shankar in
applying RG techniques to the interacting fermion
problem~\cite{FLTandRG1,Shankar}.

RG methods used to solve
multi-channel problems have several advantages. First, all channels
and the corresponding momentum scales can be treated on equal
footing in a systematic way, which can accommodate fundamental
symmetries, such as the Pauli principle. In this study, we
considered the RG flow in the particle-hole channels at the
one-loop level. An expansion in momentum transfers was performed,
which enabled us to compute the beta functions analytically. We
note that it is fairly straightforward to include the
particle-particle channel in the RG equations for the scattering
amplitude and the quasiparticle interaction. This opens the
possibility to explore the interference of the particle-hole and
particle-particle channels. A calculation of this kind corresponds
to solving the ``parquet equations at the one-loop order''. The
expansion in momentum transfers can be removed by computing the
complete particle-hole phase space~\cite{BF}.

In contrast to the traditional approach to the many-body problem,
the RG adapts the starting vacuum interaction to include the in
medium effects due to the fastest particle-hole excitations. In the
subsequent iteration, this renormalized interaction is modified to
account for the next fastest modes. Therefore, at every step, the
flow of the effective interaction is expanded around the current
system, which highlights the efficacy of the approach. Moreover,
the RG flow is physically very transparent, since the effects of
the high momentum modes are readily tractable. We believe that the
RG method is a promising tool for studying a wide range of nuclear
many-body problems.

At the one-loop level and at zero temperature the RG equations are
particularly transparent. We found an ambiguous long-wavelength
limit, analogous to that appearing in the original treatment of
Landau, which can be used to distinguish between the quasiparticle
interaction and the forward scattering amplitude. This implies
that, at the one-loop level, it is necessary to solve only the RG
equation for the scattering amplitude. The quasiparticle
interaction is then obtained in the corresponding limit. A test for
the consistency of the solution is provided by general relations
between the forward scattering amplitude and the Fermi liquid
parameters.

We solved the RG equations in the particle-hole channels for
neutron matter, where tensor interactions are absent in relative
S-states. The RG flow includes the exchange channel, and thus it
builds up the polarization effects of the induced interaction of
Babu and Brown. Our results for neutron matter are very encouraging
and provide strong motivation for further developments of the RG
approach to the nuclear many-body problem as well as for condensed
matter systems.

In the calculation, polarization and self-energy effects have been
treated self-consistently. The latter have been included in a
simple approximation, which nevertheless led to good quantitative
results at intermediate densities.

Results for the Fermi liquid parameters and the scattering
amplitude for the density range of interest to neutron stars were
presented. The Fermi liquid parameters satisfy the Pauli principle
sum rules by construction. Moreover, polarization effects lead to an
enhancement in the pressure and the energy per neutron below
nuclear matter density. As a first application, which probes the
momentum dependence of the scattering amplitude, we computed
the $^1$S$_0$ superfluid gap using weak coupling BCS theory.
A reduction from a direct gap of $3.3 \, \text{MeV}$ to
$0.8 \,\text{MeV}$ was found. Our results confirm the importance
of particle-hole polarization effects for superfluidity. A further
application of the scattering amplitude at finite momentum transfers
is in calculations of transport processes.

For a similar analysis of nuclear matter, it is necessary to extend
the flow equations to incorporate tensor interactions. These are
important for a complete discussion of nuclear matter, and we expect
rather large renormalization effects from the arguments given
in~\cite{IIpaper}. Moreover, tensor correlations and the tensor
Fermi liquid parameters play an important role in the physics of
dense matter~\cite{Olsson}. As noted in the introduction, tensor
forces are less pronounced in neutron matter, since they are not
operative in relative S-states. However, for specific quantities
the tensor force may be of crucial importance.

To date, polarization effects on the $^3$P$_2$--$^3$F$_2$ pairing
of neutrons have not been addressed and a quantitative analysis is
very much needed. Pethick and Ravenhall argue that both density and
spin-density fluctuations would increase the gap in this state and
thus possibly extend the superfluid regime to lower
densities~\cite{3pf2}. Once the RG approach is adapted to allow for
tensor forces, it should be straightforward to address this
problem~\cite{inprep}.

\begin{ack}
We thank Tom Kuo and Scott Bogner for providing us with the
$V_{\text{low k}}$ code and Kevin Bedell, Vincent Brindejonc and Janos Polonyi
for rewarding discussions. AS thanks the Theory Group at GSI for
their kind hospitality. The work of AS and GEB is supported by the
US-DOE grant No. DE-FG02-88ER40388.
\end{ack}

\appendix
\section{Derivation of the Fermi liquid theory results from the ZS
channel beta function}

For a one channel problem, such as the interacting Fermi system
restricted to the ZS channel, the one-loop RG is equivalent to solving
the Bethe-Salpeter equation in that channel. We define the fast
particle-hole Lindhard function at $\omega=0$ by integrating out the
fast modes
\begin{equation}
\chi(q,\la) = g \int\limits_{\text{fast}, \Lambda} \frac{d^3 \ip}{(2 \pi)^3} \:
\frac{n_{\ip+\q/2}-n_{\ip-\q/2}}{\varepsilon_{\ip+\q/2}
-\varepsilon_{\ip-\q/2}} .
\end{equation}
For $q \ll \kf$, we expand and write the particle-hole
phase space explicitly at zero temperature
\begin{align}
\chi(q,\la) &= g \int \frac{d^3 \ip}{(2 \pi)^3} \;
\frac{1}{v_{\text{F}} q \cos\theta} \nonumber \\[1mm]
&\times \biggl[
\Theta(\kf-\la-(p''+q \cos\theta/2)) \: \Theta((p''-q
\cos\theta/2)-(\kf+\la)) \nonumber \\[1mm]
&- \Theta((p''+q \cos\theta/2)-(\kf+\la)) \: \Theta(\kf-\la-(p''-q
\cos\theta/2)) \biggr] ,
\end{align}
where the occupation numbers are automatically taken care of by the
$\Theta$ functions that define the border between the fast and slow
modes and $\theta$ denotes the angle between $\ip$ and $\q$. We
rewrite the product of theta functions using $\Theta(\la_1-x) \:
\Theta(x-\la_2) = ( \Theta(x-\la_2) - \Theta(x-\la_1) ) \:
\Theta(\la_1-\la_2)$. This leads to
\begin{align}
\chi(q,\la) &= g \int \frac{d^3 \ip}{(2 \pi)^3} \;
\frac{1}{v_{\text{F}} q \cos\theta} \nonumber \\[1mm]
&\times \biggl[ \bigl\{ \,
\Theta(p''-q \cos\theta/2-\kf-\la) - \Theta(p''+q
\cos\theta/2-\kf+\la) \, \bigr\} \nonumber \\[1mm]
&\times \Theta(-q \cos\theta - 2 \la) - \bigl\{ \, \Theta(p''+q
\cos\theta/2-(\kf+\la)) \nonumber \\[1mm]
&- \Theta(p''-q \cos\theta/2-\kf+\la) \, \bigl\} \: \Theta(q
\cos\theta -2 \la) \biggr] .
\end{align}
When we differentiate with respect to the cutoff, the contributions
from the multiplicative theta functions $\Theta(\pm q \cos\theta - 2 \la)$
vanish, since for $q \cos\theta/2 = \pm \la$ the two remaining
theta functions cancel. Moreover, $q \ll \kf$ implies $\la \ll \kf$,
which leads to
\begin{align}
\frac{d \chi(q,\la)}{d \la} &= g \int \frac{d^3 \ip}{(2 \pi)^3} \;
\frac{1}{v_{\text{F}} q \cos\theta} \: 2 \, \delta(p''-\kf) \: \biggl[
\, \Theta(q \cos\theta - 2 \la) \nonumber \\[1mm]
&- \Theta(-2 \la - q \cos\theta) \biggr] \\[1mm]
&= \frac{2}{q} \: \frac{g \, m^\star \kf}{2 \, \pi^2} \:
\int_{2\la/q<|\cos\theta|<1} \frac{d \Omega_\ip}{4 \pi} \:
\frac{\text{sign}\bigl(\cos\theta\bigr)}{\cos\theta}
\; \Theta(q-2\la) .
\end{align}
This leads to the RG equation, Eq.~(\ref{rgeq}). Under restriction to
the ZS channel, we find after projecting the scattering
amplitude on Legendre polynomials
\begin{align}
a & = \sum_l a_l(\la) \: P_l(\cos\theta_{q'}) \\
\frac{d a_l(\la)}{d \la} &= - \frac{g \, m^\star \kf}{2 \, \pi^2}
\: \frac{2}{q} \: \log\biggl(\frac{2\la}{q}\biggr) \:
\frac{a_l(\la)^2}{2l+1} \:
\Theta(q-2\la) ,
\label{onechannelrg}
\end{align}
where we have set $z_{\kf} = 1$ for simplicity.
We propose to think of $q$ as an external scale, since it must be kept
finite and may only be set to zero after integrating the RG
equation. The solution of the RG equation in one channel,
Eq.~(\ref{onechannelrg}), is obtained by integrating from the initial
condition $a_l(\la=q/2) = a_l^{(0)}$ to $\la = 0$. We find
\begin{equation}
\displaystyle
a_l = \cfrac{a_l^{(0)}}{1+\cfrac{g \, m^\star \kf}{2 \,
\pi^2} \cfrac{a_l^{(0)}}{2l+1}} \, ,
\end{equation}
which are the standard Fermi liquid theory relations among the
quasiparticle interaction and the forward scattering amplitude
after identifying $\displaystyle A_l = \frac{g \, m^\star \kf}{2 \,
\pi^2} \, a_l$ and $\displaystyle F_l = \frac{g \, m^\star \kf}{2 \,
\pi^2} \, a_l^{(0)}$. The RG initial condition being the Fermi
liquid parameters is understood, since we assume that the flow in the
ZS' channel has been carried out to build up the quasiparticle
interaction or equivalently the Fermi liquid parameters are
taken from experimental measurements.

\end{document}